\documentclass[twocolumn]{aastex7}

\usepackage{natbib} 
\usepackage{booktabs} 

\begin{document}

\title{High-Precision Near-Infrared Abundances of Solar Analogs in the $YJ$ Bands}

\correspondingauthor{Noriyuki Matsunaga}
\email[show]{matsunaga@astron.s.u-tokyo.ac.jp}

\author[0009-0003-6827-6396,gname=Noriyuki, sname=Matsunaga]{Noriyuki Matsunaga}
\affiliation{Department of Astronomy, School of Science, The University of Tokyo, 7-3-1 Hongo, Bunkyo-ku, Tokyo 113-0033, Japan}
\affiliation{Laboratory of Infrared High-resolution spectroscopy (LiH), Koyama Astronomical Observatory, Kyoto Sangyo University, Motoyama, Kamigamo, Kita-ku, Kyoto, 603-8555, Japan}
\email{matsunaga@astron.s.u-tokyo.ac.jp}
\author[0000-0002-9397-3658,gname=Takuji,sname=Tsujimoto]{Takuji Tsujimoto}
\affiliation{National Astronomical Observatory of Japan, 2-21-1 Osawa, Mitaka, Tokyo 181-8588, Japan}
\email{taku.tsujimoto@nao.ac.jp}
\author[0000-0002-2861-4069,gname=Daisuke,sname=Taniguchi]{Daisuke Taniguchi}
\affiliation{Department of Physics, Tokyo Metropolitan University, 1-1 Minami-Osawa, Hachioji, Tokyo 192-0397, Japan}
\affiliation{National Astronomical Observatory of Japan, 2-21-1 Osawa, Mitaka, Tokyo 181-8588, Japan}
\affiliation{Laboratory of Infrared High-resolution spectroscopy (LiH), Koyama Astronomical Observatory, Kyoto Sangyo University, Motoyama, Kamigamo, Kita-ku, Kyoto, 603-8555, Japan}
\email{d.taniguchi.astro@gmail.com}
\author[0000-0001-6401-723X,gname=Hiroaki, sname=Sameshima]{Hiroaki Sameshima}
\affiliation{Institute of Astronomy, Graduate School of Science, The University of Tokyo, 2-21-1 Osawa, Mitaka, Tokyo 181-0015, Japan}
\email{hsameshima@ioa.s.u-tokyo.ac.jp}
\author[gname=Shogo, sname=Otsubo]{Shogo Otsubo}
\affiliation{Laboratory of Infrared High-resolution spectroscopy (LiH), Koyama Astronomical Observatory, Kyoto Sangyo University, Motoyama, Kamigamo, Kita-ku, Kyoto, 603-8555, Japan}
\email{otsubo@cc.kyoto-su.ac.jp}
\author[gname=Tomomi, sname=Takeuchi]{Tomomi Takeuchi}
\affiliation{Laboratory of Infrared High-resolution spectroscopy (LiH), Koyama Astronomical Observatory, Kyoto Sangyo University, Motoyama, Kamigamo, Kita-ku, Kyoto, 603-8555, Japan}
\email{takeuchi.t@cc.kyoto-su.ac.jp}
\author[gname=Yuki, sname=Sarugaku]{Yuki Sarugaku}
\affiliation{Laboratory of Infrared High-resolution spectroscopy (LiH), Koyama Astronomical Observatory, Kyoto Sangyo University, Motoyama, Kamigamo, Kita-ku, Kyoto, 603-8555, Japan}
\email{sarugaku@cc.kyoto-su.ac.jp}
\author[0009-0000-2403-9442,gname=Ilaria, sname=Petralia]{Ilaria Petralia}
\affiliation{Instituto de Astrof\'isica, Departamento de F\'isica y Astronom\'ia, Facultad de Ciencias Exactas, Universidad Andr\'es Bello, Fern\'andez Concha 700, Las Condes, Santiago, Chile}
\email{ilariapetralia28@gmail.com}
\author[0000-0001-5642-2569,gname=Scarlet,sname=Elgueta]{Scarlet Elgueta}
\affiliation{Instituto de Astrof\'isica, Departamento de F\'isica y Astronom\'ia, Facultad de Ciencias Exactas, Universidad Andr\'es Bello, Autopista Concepci\'on-Talcahuano 7100, Talcahuano, Chile}
\affiliation{Departamento de F\'isica, Universidad de Santiago de Chile, Av.~Victor Jara 3659, Santiago, Chile}
\email{sselgueta@gmail.com}
\author[0009-0000-5110-9805,gname=Matilde,sname=Coello-Guzman]{Matilde Coello-Guzm\'an}
\affiliation{Departamento de F\'isica, Universidad de Santiago de Chile, Av.~Victor Jara 3659, Santiago, Chile}
\email{matilde.coello@usach.cl}
\author[gname=Kei,sname=Fukue]{Kei Fukue}
\affiliation{Laboratory of Infrared High-resolution spectroscopy (LiH), Koyama Astronomical Observatory, Kyoto Sangyo University, Motoyama, Kamigamo, Kita-ku, Kyoto, 603-8555, Japan}
\affiliation{Education Center for Medicine and Nursing, Shiga University of Medical Science, Seta Tsukinowa-cho, Otsu, Shiga 520-2192, Japan}
\email{k.fukue@cc.kyoto-su.ac.jp}
\author[0000-0003-2380-8582,gname=Yuji,sname=Ikeda]{Yuji Ikeda}
\affiliation{Photocoding, 460-102 Iwakura-Nakamachi, Sakyo-ku, Kyoto 606-0025, Japan}
\affiliation{Laboratory of Infrared High-resolution spectroscopy (LiH), Koyama Astronomical Observatory, Kyoto Sangyo University, Motoyama, Kamigamo, Kita-ku, Kyoto, 603-8555, Japan}
\email{ikeda@photocoding.com}
\author[gname=Hideyo, sname=Kawakita]{Hideyo Kawakita}
\affiliation{Laboratory of Infrared High-resolution spectroscopy (LiH), Koyama Astronomical Observatory, Kyoto Sangyo University, Motoyama, Kamigamo, Kita-ku, Kyoto, 603-8555, Japan}
\affiliation{Department of Astrophysics and Atmospheric Sciences, Faculty of Science, Kyoto Sangyo University, Motoyama, Kamigamo, Kita-ku, Kyoto 603-8555, Japan}
\email{kawakthd@cc.kyoto-su.ac.jp}
\author[0000-0002-2662-3762,gname=Valentina,sname=D'Orazi]{Valentina D'Orazi}
\affiliation{Dipartimento di Fisica, Universita di Roma Tor Vergata, via della Ricerca Scientifica 1, 00133 Roma, Italy}
\affiliation{INAF Osservatorio Astronomico di Roma, via Frascati 33, 00078 Monte Porzio Catone, Italy}
\email{vdorazi@roma2.infn.it}
\author[0000-0002-4896-8841,gname=Giuseppe,sname=Bono]{Giuseppe Bono}
\affiliation{Dipartimento di Fisica, Universita di Roma Tor Vergata, via della Ricerca Scientifica 1, 00133 Roma, Italy}
\affiliation{INAF Osservatorio Astronomico di Roma, via Frascati 33, 00078 Monte Porzio Catone, Italy}
\email{Giuseppe.Bono@roma2.infn.it}

\begin{abstract}
We present a near-infrared abundance analysis of 46 solar analogs with known ages,
observed with the WINERED WIDE-mode spectrograph at a resolution of
$\lambda/\Delta\lambda = 28,000$. Using an empirically calibrated, line-by-line approach
in the $YJ$ bands (0.976--1.089 and 1.182--1.319~{$\mu$}m), we derive abundances for 16 elements.
Despite the intrinsic weakness of near-infrared phosphorus diagnostics, the
combination of five P\,{\sc i} lines yields a typical uncertainty half-width of
$\sim$0.04~dex, providing an estimate of the internal precision over the solar-analog sample.
For other elements, the internal precision ranges from $\sim$0.01~dex for Fe and Si
to over 0.05--0.14~dex for elements with only a couple of lines available.
The resulting per-object abundances for various elements are consistent
with previous measurements using high-precision optical spectra with
residuals of 0.03--0.2~dex depending on the element.
The inferred age--[X/Fe] relations reproduce known trends for the thin disk,
while extending them to elements difficult to access in the optical, including P and K.
We find the slope of the age--[P/Fe] relation is steeper than that for
$\alpha$ elements, which provides an empirical constraint for future modeling
of Galactic phosphorus evolution.
In addition, we publish a high signal-to-noise (S/N 500--1000) reference spectrum
constructed by combining solar-analog spectra, together with the spectra of individual stars,
and an empirically calibrated line list with per-line zero-point corrections,
for future near-infrared spectroscopic studies.
\end{abstract}

\keywords{\uat{Galaxy chemical evolution}{580} --- \uat{Solar analogs}{1941} --- \uat{Stellar abundances}{1577} --- \uat{Spectral line lists}{2082} --- \uat{Infrared spectroscopy}{2285}}


\section{Introduction} \label{sec:introduction}

High-precision abundance studies of Sun-like stars have revealed that
their photospheres encode subtle, element-by-element signatures of
Galactic chemical evolution (GCE) and stellar birth environments.
In particular, optical analyses based on line-by-line differential techniques have
demonstrated remarkably small internal star-to-star scatter
and coherent abundance patterns that correlate with stellar age across many elements,
establishing tight [X/Fe]--age relations on a highly homogeneous abundance
scale \citep[e.g.,][]{Bedell-2018,Spina-2018}\footnote{The sample in \citet{Spina-2018,Bedell-2018} was referred to as ``solar twins''
following a common usage scheme \citep{Ramirez-2009}. 
However, these stars span a wide range of ages and exhibit systematic,
age-dependent abundance variations. To emphasize their role as tracers of
Galactic chemical evolution rather than identical counterparts of the Sun,
we adopt the term ``solar analogs'' throughout this paper.}.
Interpreting these empirical trends requires not only nucleosynthetic yields but
also a dynamical framework for the Galactic disk: \citet{Tsujimoto-2021} showed that
the observed age dependence of abundance patterns can be explained by radial migration,
linking local [X/Fe]--age sequences to stars formed at different Galactocentric radii
and subsequently mixed into the solar neighborhood.
More recently, the abundance patterns first established in small, high-precision
samples have also been explored in much larger solar-analog catalogs,
including a GALAH-based sample that demonstrates the feasibility of population-scale
high-precision work and a Gaia-based sample in which several age--[X/Fe] relations
were statistically confirmed \citep{Walsen-2024,Taniguchi-2026}.


Previous studies of solar analogs
in the optical provide a crucial reference frame for cross-calibration and
differential analysis \citep{Bedell-2018,Spina-2018}, although 
they leave some important gaps. For instance, 
optical stellar spectra of long-lived solar-type or later-type stars
lack measurable phosphorus absorption lines.
OB-type stars show lines of ionized phosphorus in the optical
\citep{Takeda-2024,Aschenbrenner-2025},
but their short lifetimes prevent tracing GCE over long timescales.
Instead, late-type stars show neutral phosphorus lines in the near-infrared,
in the $Y$ band (around 1\,{$\mu$}m) and the $H$ band (around 1.6\,{$\mu$}m).
These near-infrared P\,{\sc i} lines have enabled phosphorus abundance measurements
in several hundred stars \citep[e.g.,][]{Kato-1996,Caffau-2011,Caffau-2016,Caffau-2019,Maas-2022,Nandakumar-2022,Jian-2026}.
Neutral phosphorus lines are also found in the ultraviolet, $\lambda \lesssim 2500$\,{\AA}\ \citep{Jacobson-2014,Roederer-2014},
but previous observations in this wavelength range are 
insufficient to characterize relative abundance and temporal evolution well.

Massive stars are thought to be the main producers of phosphorus, synthesizing
$^{31}$P during advanced burning stages prior to the explosion of core-collapse supernovae (CCSNe).
In many GCE calculations, CCSNe provide the
baseline P production but standard yield sets tend to underproduce the observed disk
abundances unless the CCSN P yields are increased artificially by a factor of {$\sim$}3
\citep[e.g.,][]{Cescutti-2012,Kobayashi-2020,Nandakumar-2022}.
Type Ia supernovae (SNe Ia), while essential for Fe enrichment, are generally expected to contribute
negligibly to P in commonly-used yield sets \citep[e.g.,][]{Iwamoto-1999,Cescutti-2012}.
AGB stars can synthesize $^{31}$P via neutron-capture channels (linked to the Si isotopes),
but published yield grids indicate that their net P contribution is too small to account for the
Galactic P budget on its own \citep[e.g.,][]{Karakas-2014,Karakas-2016,Kobayashi-2020}.
Additional production in ONe novae has also been proposed,
where thermonuclear runaways can drive nucleosynthesis into the Si--P mass region,
but the predicted yields are sensitive to uncertain key reaction rates
\citep[e.g.,][]{Jose-1998,Jose-2001,Bekki-2024,Kemp-2024}.
Previous solar-analog studies have discussed the relative roles of such nucleosynthetic sources
for many elements, but phosphorus has been absent from the high-precision
solar-analog age--abundance framework.
While \citet{Melendez-2009} measured P in a small sample
(summarized by \citealt{Caffau-2011}), the sample size and the lack of homogeneous age-resolved analyses
have so far prevented direct constraints on [P/Fe] as a function of age.

In this paper, we present abundance measurements, including phosphorus, for 46 solar analogs
drawn from the sample of \citet{Bedell-2018} and
observed with the near-infrared spectrograph WINERED covering the $YJ$ bands.
Technically, solar analogs are excellent calibrators, and we use them to
establish a precise near-infrared abundance scale.
In particular, stars with ages of $\sim$3--7~Gyr exhibit nearly constant [X/Fe] patterns 
across many elements \citep{Tsujimoto-2021},
suggesting that abundance ratios for phosphorus and other previously unmeasured species
are also close to solar in this age range. We therefore use these solar analogs to calibrate 
and validate $YJ$-band line diagnostics, and we establish a consistent near-infrared abundance
scale that enables direct comparison to optical [X/Fe]--age relations.
We also provide a high-S/N combined reference spectrum together with spectra of individual stars
and an empirically corrected line list for community use.

\section{Observations and Spectral Reduction} \label{sec:obs}

\subsection{Observations} \label{subsec:obs}
We collected $YJ$-band spectra of 46 solar analogs using the WINERED spectrograph
mounted on the Magellan Clay telescope, Las Campanas Observatory, Chile \citep{Ikeda-2022,Otsubo-2024}.
All observations were carried out in the WIDE mode, providing a resolution of
$R\equiv \lambda/\Delta\lambda = 28{,}000$.
The data were obtained during 
three observing runs in 2025 February, August, and October (Table~\ref{tab:targets}).
In addition to the scientific targets, we use a library of 
A0V standard stars spanning multiple observing runs to construct
a PCA-based model of telluric absorption.
These standards were observed during multiple observing runs between 2017 and 2025.
WINERED was mounted on the New Technology Telescope (NTT) in 2017--2018 \citep{Ikeda-2018,Ikeda-2022},
and has been operated at the Magellan Clay telescope since 2022 \citep{Otsubo-2024}.

\tabletypesize{\scriptsize}
\begin{deluxetable}{lcrrrc}
\tablecaption{Targets and Observations\label{tab:targets}}
\tablehead{
\colhead{Star} & \colhead{Date (UTC)} & \colhead{$t_{\mathrm{exp}}$} & \colhead{$v_{\mathrm{broad}}$} & \colhead{S/N} & \colhead{c} 
}
\startdata
HIP\,54102 & 2025-02-03 06:43 & 40~(4) & 12.50 & 220 &  \\
HIP\,64673 & 2025-02-03 06:50 & 40~(4) & 13.11 & 259 & c \\
HIP\,41317 & 2025-02-03 07:12 & 60~(10) & 12.66 & 313 &  \\
HIP\,43297 & 2025-02-03 07:25 & 36~(6) & 12.51 & 259 &  \\
HIP\,49756 & 2025-02-03 07:35 & 24~(4) & 11.93 & 317 & c \\
HIP\,54582 & 2025-02-03 07:41 & 10~(5) & 12.58 & 272 & c \\
HIP\,54287 & 2025-02-10 06:59 & 16~(4) & 12.32 & 257 & c \\
HIP\,44713 & 2025-02-10 07:05 & 16~(4) & 12.31 & 159 &  \\
HIP\,36515 & 2025-02-10 07:15 & 15~(5) & 12.84 & 172 &  \\
HIP\,30037 & 2025-02-11 05:00 & 180~(6) & 12.24 & 274 & c \\
HIP\,38072 & 2025-02-11 05:10 & 180~(4) & 13.33 & 349 &  \\
HIP\,30502 & 2025-02-11 05:19 & 120~(4) & 12.31 & 351 & c \\
HIP\,30158 & 2025-02-11 05:26 & 80~(4) & 12.18 & 301 &  \\
HIP\,44935 & 2025-02-11 05:41 & 80~(4) & 12.41 & 324 & c \\
HIP\,44997 & 2025-02-11 05:56 & 80~(4) & 12.07 & 360 & c \\
HIP\,34511 & 2025-02-11 06:10 & 120~(6) & 12.81 & 314 & c \\
HIP\,62039 & 2025-02-11 06:23 & 40~(4) & 12.17 & 377 & c \\
HIP\,65708 & 2025-02-11 06:32 & 40~(4) & 11.85 & 391 &  \\
HIP\,74389 & 2025-02-11 06:40 & 40~(4) & 12.76 & 271 & c \\
HIP\,102152 & 2025-08-05 03:56 & 120~(4) & 12.57 & 257 &  \\
HIP\,114615 & 2025-08-05 04:11 & 360~(6) & 12.85 & 350 &  \\
HIP\,114328 & 2025-08-05 04:23 & 120~(4) & 13.11 & 285 & c \\
HIP\,96160 & 2025-08-05 05:12 & 120~(4) & 13.20 & 286 &  \\
HIP\,104045 & 2025-08-05 05:28 & 60~(4) & 12.86 & 210 & c \\
HIP\,3203 & 2025-08-05 05:39 & 20~(4) & 13.46 & 144 &  \\
HIP\,1954 & 2025-08-05 05:52 & 40~(4) & 12.53 & 398 & c \\
HIP\,115577 & 2025-08-14 08:49 & 40~(4) & 12.78 & 337 &  \\
HIP\,118115 & 2025-08-14 08:55 & 60~(4) & 13.57 & 429 &  \\
HIP\,116906 & 2025-08-14 09:00 & 60~(4) & 12.99 & 510 & c \\
HIP\,117367 & 2025-08-14 09:06 & 60~(4) & 12.80 & 482 & c \\
HIP\,10175 & 2025-08-14 09:14 & 80~(4) & 12.71 & 420 & c \\
HIP\,9349 & 2025-08-14 09:21 & 60~(4) & 12.64 & 340 &  \\
HIP\,10303 & 2025-08-14 09:25 & 40~(4) & 12.71 & 330 & c \\
HIP\,14614 & 2025-08-14 09:31 & 60~(4) & 12.73 & 378 & c \\
HIP\,15527 & 2025-08-14 09:41 & 60~(4) & 12.80 & 412 &  \\
HIP\,14501 & 2025-08-14 09:47 & 60~(6) & 12.80 & 665 &  \\
HIP\,101905 & 2025-10-01 04:47 & 20~(4) & 12.81 & 204 &  \\
HIP\,108158 & 2025-10-01 04:55 & 30~(6) & 12.63 & 217 &  \\
HIP\,105184 & 2025-10-01 05:02 & 12~(4) & 12.60 & 378 &  \\
HIP\,18844 & 2025-10-01 05:10 & 12~(4) & 12.50 & 273 & c \\
HIP\,109821 & 2025-10-01 05:17 & 8~(4) & 12.56 & 421 &  \\
HIP\,4909 & 2025-10-01 05:26 & 80~(4) & 13.37 & 387 &  \\
HIP\,5301 & 2025-10-01 05:33 & 60~(4) & 12.49 & 378 &  \\
HIP\,6407 & 2025-10-01 05:42 & 80~(4) & 12.70 & 306 &  \\
HIP\,8507 & 2025-10-01 05:49 & 60~(4) & 12.41 & 312 & c \\
HIP\,7585 & 2025-10-01 05:58 & 12~(4) & 12.88 & 397 & c \\
\enddata
\tablecomments{Date (UTC) gives the calendar date and mid-exposure time (HH:MM) in UTC. $t_{\mathrm{exp}}$ is the total exposure time in seconds with the number of exposures in parentheses. $v_{\mathrm{broad}}$ is the Gaussian broadening velocity including the instrumental width in km~s$^{-1}$. S/N per pixel is measured in order m55 of the WINERED spectrum. The final column flags calibrators (3--7~Gyr) with ``c''.}
\end{deluxetable}

\subsection{Spectral Reduction} \label{subsec:reduction}
The initial spectral reduction was performed using 
the WINERED Automatic Reduction Pipeline \citep[WARP;][]{Hamano-2024},
which produces wavelength-calibrated one-dimensional spectra from the 2D echellogram images.
Telluric absorption lines in continuum-normalized spectra were removed using
the TerraPCA package (Appendix~\ref{sec:telluric}),
implementing an approach based on the principal component analysis (PCA).
TerraPCA is designed to fit telluric absorption directly to science spectra that
contain both stellar and telluric lines, but for the solar analogs we first removed
stellar lines using the combined solar-analog spectrum (Section~\ref{subsec:combined_spec})
as the stellar template.
We divided 1D spectra from individual exposures by the combined solar-analog spectrum,
which was shifted to match the stellar radial velocity of each exposure,
and then fitted the telluric model to these stellar-line-removed spectra
in order to improve the precision and robustness of the telluric modeling.
We fitted the telluric absorption model to individual exposures
and then combined the telluric-corrected spectra for each target star.
For orders m53 and m54 (10295--10685\,{\AA}), where telluric absorption is negligible, 
we used the combined spectra without telluric correction.
Doppler shifts of the combined spectra were measured and removed
to place stellar absorption lines at their rest wavelengths.
Throughout this work, we adopt air wavelengths.

In combining spectra from individual exposures for each star,
we computed the RMS (root-mean-square)
of differences between exposures to estimate the per-pixel uncertainties.
These uncertainties can be converted to S/N (signal-to-noise ratio) 
values per pixel, approximately corresponding to half a resolution element.
The S/N (median of per-pixel flux divided by error) in the m55 order,
around 1.02~{$\mu$}m, are listed in Table~\ref{tab:targets},
typically around 250--350 and $\sim$150 in the worst cases.
Solar analogs show relatively flat spectral energy distributions,
and the S/N values are generally uniform across the $YJ$ bands.
We focus on the wavelength ranges of 9760--10890\,{\AA}\ ($Y$ band) and 11820--13190\,{\AA}\ ($J$ band),
where the telluric absorption is relatively weak.
The final science-ready spectra are available online.
\footnote{The reduced spectra are available via Zenodo (DOI: 10.5281/zenodo.19230423).}

\subsection{Combined Solar-Analog Spectrum} \label{subsec:combined_spec}
Taking advantage of the homogeneous nature of the solar-analog sample,
we also constructed a combined spectrum of 22 calibrator stars with ages
between 3 and 7~Gyr (Table~\ref{tab:targets}).
Because these stars exhibit very similar abundance patterns \citep{Bedell-2018,Tsujimoto-2021}, 
their intrinsic spectra are expected to be nearly identical.
The combined spectrum achieves significantly higher S/N than individual spectra
not only because statistical noise was reduced, but also because 
residual telluric features and other spurious noise were efficiently removed
as outliers ($\pm 3\sigma$ clipping, two iterations) during the combination.
Doppler shifts of the stellar spectra are different, while telluric absorption lines
remain at rest in the observatory frame but not aligned after the Doppler correction in the stellar frame.
The resulting average spectrum provides a useful reference for solar-analog spectra
in this wavelength range; 
it has per-pixel S/N around 1000 (standard error as low as 0.001), while the RMS of 
the differences between individual calibrators' spectra is around 0.005 in the continuum.
The combined spectrum also serves as a diagnostic tool to identify spurious noise  
or artifacts in individual solar-analog spectra, which are expected to approximately match the combined spectrum.
We define the upper and lower limits as the combined spectrum $\pm 3\times$RMS
and use them in Section~\ref{subsec:abundances} to identify and mask outlier pixels.
The combined spectrum is presented in Appendix~\ref{sec:spectra_with_lines}
and is available online via Zenodo (DOI: 10.5281/zenodo.19230423),
together with spectra of individual stars.

\section{Methods} \label{sec:abundance}

\subsection{Overview} \label{subsec:overview}

The abundance analysis consists of two stages:
(1)~selection and empirical calibration of absorption lines, and
(2)~abundance measurements for individual stars.
Both stages are based on ``residual curves,'' which quantify the mismatch between
observed and synthetic spectra as a function of trial abundance for each line.
For a given line, residual curves from different calibrator stars are combined
to determine the line-specific abundance zero-point and its scatter,
whereas for a given target star, residual curves from multiple accepted lines
are combined to infer the elemental abundance and its uncertainty-like interval.
Throughout the analysis, each ionization stage is treated independently
(e.g., Fe\,{\sc i} and Fe\,{\sc ii} are not combined).

Section~\ref{subsec:synthesis} summarizes the spectral synthesis setup and adopted stellar parameters,
and Section~\ref{subsec:rms_curve} defines the residual curves and their normalization.
Section~\ref{subsec:calibration} describes the line selection and empirical calibration based on
22 solar analogs with ages of 3--7~Gyr, while Section~\ref{subsec:estimation}
explains how the calibrated residual curves are combined to derive final abundances.

\subsection{Spectral Synthesis and Stellar Parameters} \label{subsec:synthesis}

For the spectral synthesis in this study, we use
the OCTOMAN (Optimization Code To Obtain Metallicity using Absorption liNes)
developed by \citet{Taniguchi-2025}. OCTOMAN is a wrapper
of the spectral synthesis code MOOG \citep{Sneden-1973} with additional
automation and optimization functions.
MOOG\footnote{We used the 2019 November version of MOOG with minor modifications by
M.~Jian (\url{https://github.com/MingjieJian/moog_nosm}).}
solves the 1D radiative transfer under the assumption of local thermodynamic equilibrium (LTE).
We adopt the ATLAS9-APOGEE atmosphere models \citep{Meszaros-2012}.
As the internal solar reference for the synthesis and related calculations,
we adopt the abundance compilation of \citet{Asplund-2009};
however, as described in Section~\ref{subsec:calibration}, the final abundances are placed on an
empirical line-by-line scale through calibration with solar analogs.

The stellar parameters required for the synthesis are adopted from
\citet{Spina-2018}, who determined them from a differential analysis of optical spectra
relative to the Sun.
We use these literature values directly for $T_\mathrm{eff}$, $\log g$, [Fe/H],
microturbulent velocity, and stellar age.
The only parameter newly determined in this work is the line broadening parameter
$v_\mathrm{broad}$, which includes instrumental, macroturbulent, and rotational broadening.
We estimated $v_\mathrm{broad}$ for each star by fitting synthetic spectra to isolated
Fe\,{\sc i}, Si\,{\sc i}, and Mg\,{\sc i} lines in the WINERED spectra.
The resulting values are listed in Table~\ref{tab:targets}.

\subsection{Construction and Normalization of Residual Curves} \label{subsec:rms_curve}

For each combination of star, absorption line, and line list (VALD or MB99),
we synthesize spectra over a grid of trial abundances
from $-1.5$ to $+1.5$~dex in steps of 0.1~dex, and evaluate the mismatch between
the observed and synthetic spectra within a window of $\pm 30$~km~s$^{-1}$
around the line center.
For each star--line pair, we define the residual curve as a function of trial abundances [X/H]:
\begin{equation}
  S(\mathrm{[X/H]}) = \frac{1}{N_\mathrm{pix}} \sum_{i=1}^{N_\mathrm{pix}}
    \left\{ F_\mathrm{obs}(\lambda_i) - F_\mathrm{syn}(\lambda_i, \mathrm{[X/H]}) \right\}^2,
\end{equation}
where $F_\mathrm{obs}$ and $F_\mathrm{syn}$ are the observed and synthetic fluxes,
and $N_\mathrm{pix}$ is the number of pixels within the line window.
In these calculations, the continuum level is the only free parameter;
we optimize $F_\mathrm{syn}$ by adding a constant to minimize the residual at each trial abundance.
The wavelength offset is fixed to zero and the line broadening is fixed
to the adopted $v_\mathrm{broad}$ for each star.

\begin{figure}[!tbp]
\centering
\includegraphics[width=0.90\linewidth]{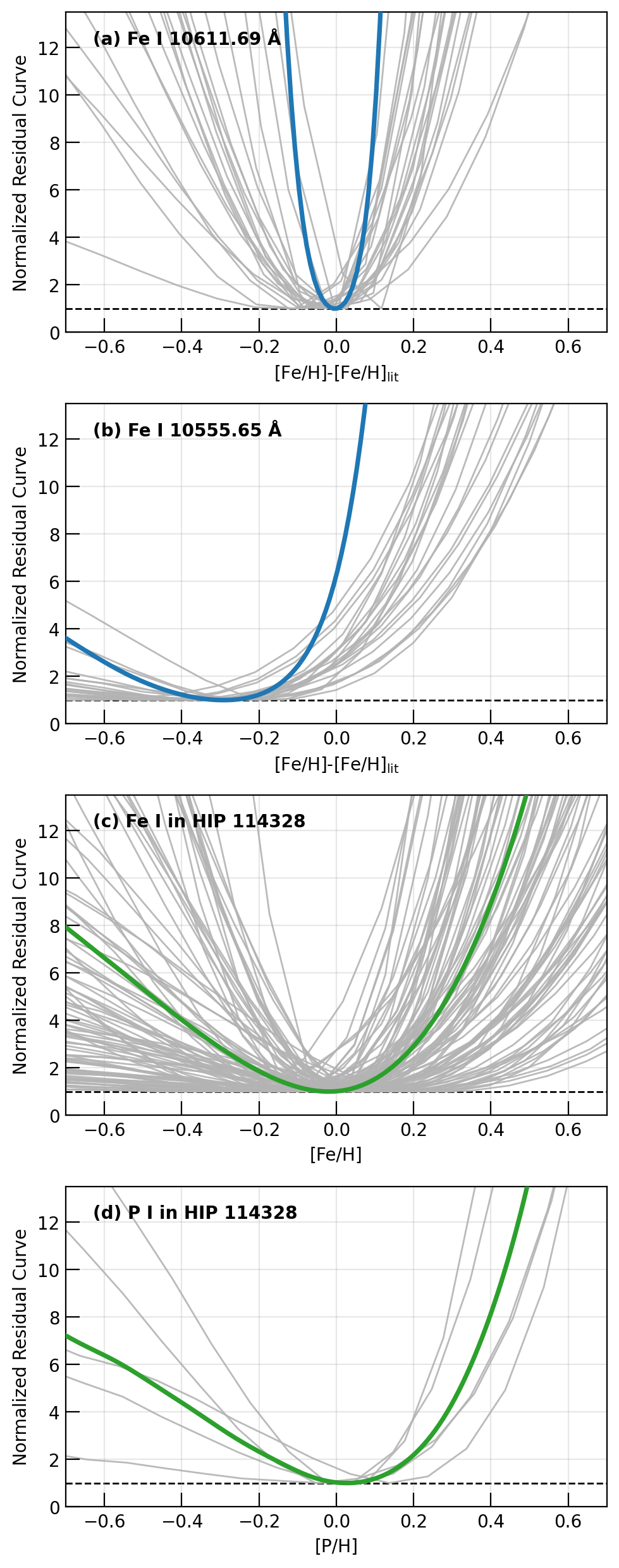}
\caption{Examples of residual curves.
Panels (a,b) show the curves for Fe~I~10611.69\,\AA\ and Fe~I~10555.65\,\AA\ measured in solar-analog calibrators.
Panels (c,d) show the target-stage combined residual curves for HIP\,114328:
Fe\,{\sc i} in (c) and P\,{\sc i} in (d).
Gray curves are individual residual curves,
while the thick blue/green curves are the combined residual curves.
All curves are normalized to their minima for illustration.}
\label{fig:rms_curve_examples}
\end{figure}

A sufficiently significant line generally yields a residual curve with a well-defined minimum,
providing a \textit{two-sided} abundance constraint.
Very weak lines show a flat curve on the low-abundance side but a steep rise on the high-abundance side,
corresponding to an upper limit;
such lines can still contribute useful information when combined with other lines.
Figure~\ref{fig:rms_curve_examples}(a,b) shows examples for strong and weak Fe\,{\sc i} lines.
For clarity, the plotted curves are normalized to their minima.

We normalize the residual curves using an effective number of degrees of freedom (DOF),
for which we take into account correlations between neighboring pixels in the spectra. 
The DOF is estimated from the effective number
of independent pixels $N_\mathrm{eff}$ (Appendix~\ref{sec:effective_dof}):
\begin{equation}
  \mathrm{DOF} = \left(\frac{N_{\mathrm{eff}}}{N}\right) N_{\mathrm{pix}} - N_{\mathrm{par}},
\end{equation}
where $N_{\mathrm{par}}$ accounts for the abundance and continuum level.
We adopt $N_{\mathrm{eff}}/N \simeq 0.25$, which gives DOF values of order unity
for the present line windows.
In the following analysis, residual curves are normalized either to the effective DOF (or to unity) at the minimum,
depending on the context, and the corresponding uncertainty-like interval is defined by the
standard ``DOF$+1$'' (or $1+1/\mathrm{DOF}$) criterion, respectively.

\subsection{Line Selection and Empirical Calibration} \label{subsec:calibration}

We first construct a candidate line list by inspecting spectra of solar analogs and Cepheids
and by including lines used in our previous studies
\citep{Kondo-2019,Fukue-2021,Matsunaga-2020,Elgueta-2024}.
We consider transition data for all atoms and ions available in 
the Vienna Atomic Line Database \citep[VALD;][]{Ryabchikova-2015}
and in \citet[][hereafter MB99]{Melendez-1999}.

We then apply a depth filter, requiring that more than half of the solar analogs
show an absorption depth greater than 0.01 at the line center.
This filter removes very weak candidates before the residual-curve analysis,
but does not guarantee that the measured absorption is dominated by the transition of interest.
We retain strong lines in this work; their behavior in solar analogs is discussed in
Appendix~\ref{sec:strong_lines}.
In addition, we estimate the expected importance of blending from other elements, for each line,
using the ratio
\begin{equation}
  \beta_1 = W_\mathrm{without}/W_\mathrm{with},
\end{equation}
where $W_\mathrm{with}$ and $W_\mathrm{without}$ are the equivalent widths
computed with and without the element of interest, respectively,
within $\pm 15$~km~s$^{-1}$ of the line center for a solar model atmosphere.

For the line calibration, we consider only the 22 solar analogs with literature ages between
3 and 7~Gyr, indicated by the ``c'' flag in Table~\ref{tab:targets}, in order to minimize age-dependent abundance variations.
Previous studies \citep{Bedell-2018,Tsujimoto-2021} showed that stars in this age range
have nearly solar abundance ratios for the elements measured in the optical.
We therefore assume solar abundance ratios for elements without previous measurements
(e.g., phosphorus) when calibrating their lines.
Residual curves for the remaining stars are also measured in the same way and used
in the later abundance analysis.

For each absorption line, we combine the residual curves of the calibrator stars on a common
abundance-offset axis, $\Delta X$.
For elements measured by \citet{Spina-2018} and \citet{Bedell-2018},
we define $\Delta X = \mathrm{[X/H]} - \mathrm{[X/H]_{lit}}$, where [X/H] indicates the trial abundance;
for other elements, we assume [X/Fe]$=0$ in the calibrator sample and adopt
$\Delta X = \mathrm{[X/H]} - \mathrm{[Fe/H]_{lit}}$.
The weighted sum of these curves yields a combined residual curve for each line,
and the combined curve is normalized to unit minimum.
Outlying residual curves from individual stars are iteratively rejected if their
allowed ranges do not overlap that of the combined curve or if they are strongly displaced
from the origin.

The minima of the individual residual curves generally exhibit both statistical 
and systematic offsets from zero. The systematic offsets may arise from inaccurate $\log gf$ values
or from imperfections in spectral synthesis, such as blending lines that affect
all calibrator stars in a similar manner. The combined residual curve mentioned above thus keeps
the line systematics but shows smaller statistical errors than individual residual curves.
We estimate the {\it net} effects of these systematic errors empirically as the per-line zero-point offset,
$\Delta_\mathrm{line}$, and the uncertainty, $\sigma_\mathrm{line}$, based on
the combined residual curve.
The location of the minimum defines $\Delta_\mathrm{line}$, while
$\sigma_\mathrm{line}$ is given by the half-width of the uncertainty-like intervals,
defined as the range over which the combined curve
remains below $1+1/\Sigma_\mathrm{DOF}$, where $\Sigma_\mathrm{DOF}$ is the sum of
the DOF values of the residual curves combined.

Finally, we retain only lines that satisfy all of the following criteria:
\begin{itemize}
\item the combined residual curve provides a well-defined \textit{two-sided} constraint with
      at least 11 contributing calibrator stars;
\item $\Delta_\mathrm{line}$ lies within $\pm 0.5$~dex of the median value for the element;
\item $\sigma_\mathrm{line} \leq 0.25$~dex;
\item $\beta_1 < 0.5$.
\end{itemize}
The resulting $\Delta_\mathrm{line}$ and $\sigma_\mathrm{line}$ values are then used in the
final abundance analysis.

\subsection{Abundance Estimation from Combined Residual Curves} \label{subsec:estimation}

For each combination of star, element, and line list (VALD or MB99),
we collect all accepted residual curves for absorption lines.
Each per-line residual curve is shifted by $-\Delta_\mathrm{line}$ to remove the empirical
zero-point offset and then convolved with a Gaussian of width $\sigma_\mathrm{line}$  
to propagate the per-line scatter estimated during the calibration stage.
First, the normalized curves are interpolated onto a common [X/H] grid with
a finer step of 0.002~dex to locate the minimum precisely.
They are then summed to form the combined residual curve, whose minimum
is then renormalized to unity.
The abundance corresponding to this minimum is adopted as the best estimate of the elemental abundance [X/H],
and its error denoted $e_\mathrm{X}$ is given by the half-width of the uncertainty-like intervals
under the threshold of $1+1/\mathrm{DOF}$ as we do in line calibration (Section~\ref{subsec:calibration}).

Figure~\ref{fig:rms_curve_examples}(c,d) shows examples of such combined residual curves
for Fe\,{\sc i} and P\,{\sc i} in HIP\,114328 using the VALD line list.
Gray curves represent the per-line residual curves after applying the $\Delta_\mathrm{line}$ shifts,
while the colored curves show the weighted combination for each element.

\section{Results} \label{sec:results}

In this section, we first discuss the accepted line lists and their empirical calibration,
then present the resulting abundances of the solar analogs, and finally focus on the constraints
obtained for phosphorus.

\subsection{Line Selection and Calibration} \label{subsec:line_selection}

\begin{figure*}[!tbp]
\centering
\includegraphics[width=0.95\linewidth]{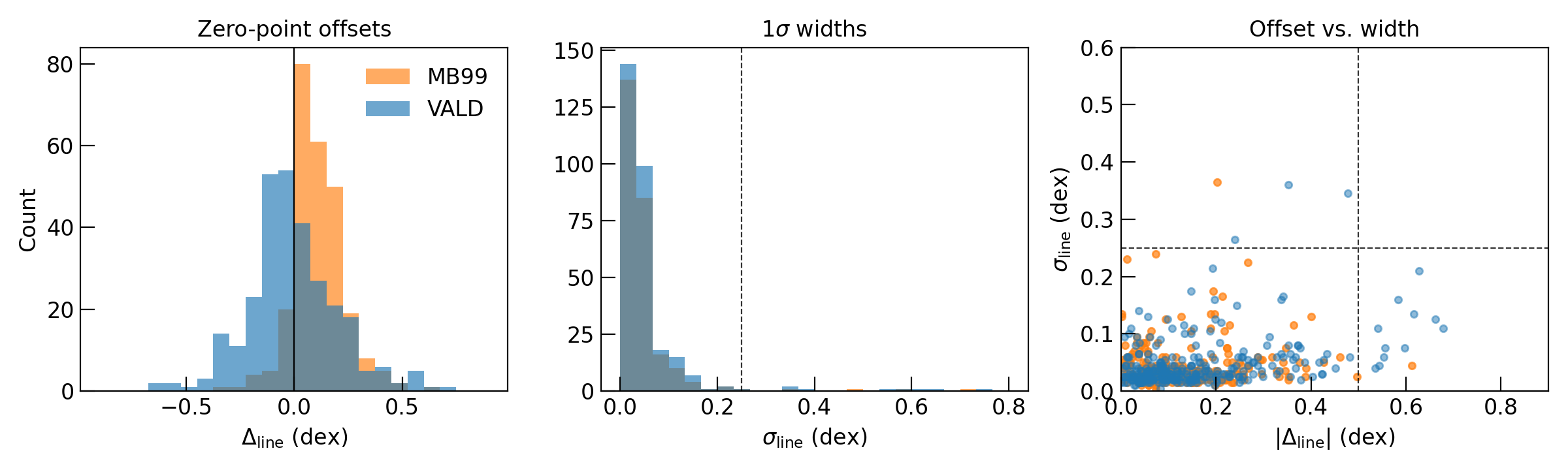}
\caption{Distributions of per-line zero-point offsets $\Delta_\mathrm{line}$
and half-widths $\sigma_\mathrm{line}$ for candidate lines taken from VALD (blue)
and those from MB99 (orange).
Left: histograms of $\Delta_\mathrm{line}$. Center:
histograms of $\sigma_\mathrm{line}$.
Right: $|\Delta_\mathrm{line}|$ versus $\sigma_\mathrm{line}$.
Most lines cluster around small offsets and narrow widths, while several lines
were rejected due to large offsets ($|\Delta_\mathrm{line}|>0.5$) or half-widths ($\sigma_\mathrm{line}>0.25$).}
\label{fig:line_cal_stats}
\end{figure*}

First, we constructed a candidate line list consisting of 398 VALD lines and 315 MB99
lines.\footnote{Throughout the analysis in this work, VALD and MB99 are treated independently.}
After the depth screening step, which requires absorption deeper than 0.01
(i.e.~normalized flux lower than 0.99) at line center in
more than half of the solar analogs, 312 VALD and 265 MB99 lines remained.  
We computed residual curves for these lines for all solar analogs.
Using only the 22 calibrator stars with ages of 3--7~Gyr,
we then constructed combined residual curves and derived the empirical per-line zero-point offsets,
$\Delta_\mathrm{line}$, and their uncertainties, $\sigma_\mathrm{line}$.  
During this calibration step, we iteratively rejected outlier residual curves whose constraints 
deviate significantly from the combined curve;
after each rejection step, the combined residual curve was recomputed and normalized.
We estimated $\Delta_\mathrm{line}$ from the location of the minimum
and $\sigma_\mathrm{line}$ as the half-width of the $\pm 1$ interval
over which the combined and normalized residual curve remains
below $1+1/\Sigma_\mathrm{DOF}$ (Section~\ref{subsec:calibration}).
Figure~\ref{fig:line_cal_stats} shows the distributions of $\Delta_\mathrm{line}$
and $\sigma_\mathrm{line}$ derived for candidate lines. For most lines, the zero-point offsets
are moderately small, within $\pm 0.2$~dex, and the $\sigma_\mathrm{line}$ values 
are also tightly clustered, mostly below 0.1~dex. 

Applying the acceptance criteria described in Section~\ref{subsec:calibration},
we finally retained 256 VALD and 237 MB99 lines for the abundance measurements.
Table~\ref{tab:accepted_lines} lists the accepted lines together with the derived $\Delta_\mathrm{line}$
and $\sigma_\mathrm{line}$ values, as well as the line depths measured in the combined spectrum.
We also assign the ``m'' flag to lines for which multiple transitions of the same species
within $\pm 30$~km~s$^{-1}$ contribute significantly to the observed absorption.
This is assigned using the $X$ line-strength indicator \citep{Magain-1984,Gratton-2006},
\begin{equation}
	X=\log gf - \mathrm{EP} \times \frac{5040}{0.86\times 5780} ,
\end{equation}
by searching for nearby lines with $X$ values within 0.3~dex of the primary transition.
We find 19 lines with the ``m'' flag, including multiplets and lines with hyperfine structure
\citep[i.e., Mn\,{\sc i} 12899.67\,{\AA}\ and 12975.94\,{\AA}; ][]{Melendez-1999b,Blackwell-Whitehead-2011}.
The numbers of accepted lines 
per element are summarized for both VALD and MB99 lines in Table~\ref{tab:lines_stats}. 
For illustration, Appendix~\ref{sec:spectra_with_lines} shows the accepted lines
overplotted on the combined spectrum of the calibrator stars (3--7~Gyr).

\begin{deluxetable}{lrrrrrrrrrrr}
\tablecaption{Accepted Lines for Solar Analogs (first 10 shown; full table machine-readable)\label{tab:accepted_lines}}
\tablehead{\colhead{Species} & \colhead{$\lambda_\mathrm{air}$ (\AA)} & \colhead{Depth} & \multicolumn{4}{c}{VALD} & \multicolumn{4}{c}{MB99} & \colhead{m} \\ 
\cmidrule(r){4-7} \cmidrule(l){8-11}
 &  &  & \colhead{EP} & \colhead{$\log gf$} & \colhead{$\Delta_\mathrm{line}$} & \colhead{$\sigma_\mathrm{line}$} & \colhead{EP} & \colhead{$\log gf$} & \colhead{$\Delta_\mathrm{line}$} & \colhead{$\sigma_\mathrm{line}$} &  }
\startdata
C~{\sc i} & 10123.87 & 0.127 & 8.537 & $-0.031$ & $+0.086$ & 0.025 & 8.54 & $-0.09$ & $+0.126$ & 0.020 & - \\
C~{\sc i} & 10541.23 & 0.031 & 8.537 & $-1.398$ & $+0.256$ & 0.060 & 8.54 & $-1.27$ & $+0.056$ & 0.065 & - \\
C~{\sc i} & 10683.09 & 0.281 & 7.483 & $0.079$ & $+0.296$ & 0.035 & 7.48 & $0.03$ & $+0.296$ & 0.030 & - \\
C~{\sc i} & 10685.36 & 0.225 & 7.480 & $-0.272$ & $+0.056$ & 0.030 & 7.48 & $-0.30$ & $+0.176$ & 0.020 & - \\
C~{\sc i} & 10691.26 & 0.328 & 7.488 & $0.344$ & $+0.356$ & 0.065 & 7.49 & $0.28$ & $+0.426$ & 0.050 & - \\
C~{\sc i} & 10707.34 & 0.211 & 7.483 & $-0.411$ & $+0.156$ & 0.025 & 7.48 & $-0.41$ & $+0.106$ & 0.025 & - \\
C~{\sc i} & 10729.54 & 0.202 & 7.488 & $-0.420$ & $+0.106$ & 0.025 & 7.49 & $-0.46$ & $+0.156$ & 0.020 & - \\
C~{\sc i} & 10753.99 & 0.059 & 7.488 & $-1.606$ & $+0.036$ & 0.040 & 7.49 & $-1.69$ & $+0.086$ & 0.030 & - \\
C~{\sc i} & 11848.73 & 0.059 & 8.643 & $-0.697$ & $+0.036$ & 0.030 & 8.64 & $-0.70$ & $-0.014$ & 0.030 & - \\
C~{\sc i} & 11862.97 & 0.056 & 8.640 & $-0.710$ & $-0.084$ & 0.050 & 8.64 & $-0.70$ & $-0.094$ & 0.050 & - \\
\enddata
	\tablecomments{EP and $\log gf$ are the excitation potential and oscillator strength (before calibration) from the original VALD and MB99 line lists. Depth is the line-center depth (1$-$mean flux) in the combined calibrator spectrum (Section~\ref{subsec:combined_spec}). $\Delta_\mathrm{line}$ is the median line-by-line abundance correction for that line list, and $\sigma_\mathrm{line}$ is the scatter of those corrections. The flag ``m'' marks lines where a multi-line contribution is expected.}
\end{deluxetable}

\begin{deluxetable}{crrrr}
\tablecaption{Number of Selected Lines and Mean Abundance Errors\label{tab:lines_stats}}
\tablehead{\colhead{Species} & \multicolumn{2}{c}{VALD} & \multicolumn{2}{c}{MB99} \\
\cmidrule(r){2-3} \cmidrule(l){4-5}
 \colhead{} & \colhead{$N_\mathrm{line}$} & \colhead{$\langle e_\mathrm{X}\rangle$} & \colhead{$N_\mathrm{line}$} & \colhead{$\langle e_\mathrm{X}\rangle$} }
\startdata
	C\,{\sc i} & 19 & $0.011\,(0.010)$ & 25 & $0.011\,(0.011)$ \\
	Na\,{\sc i} & 1 & $0.102\,(0.107)$ & 2 & $0.034\,(0.032)$ \\
	Mg\,{\sc i} & 8 & $0.012\,(0.013)$ & 6 & $0.014\,(0.016)$ \\
	Al\,{\sc i} & 3 & $0.067\,(0.066)$ & 3 & $0.067\,(0.069)$ \\
	Si\,{\sc i} & 55 & $0.005\,(0.005)$ & 52 & $0.004\,(0.004)$ \\
	P\,{\sc i} & 5 & $0.033\,(0.035)$ & 4 & $0.041\,(0.046)$ \\
	S\,{\sc i} & 3 & $0.024\,(0.025)$ & 4 & $0.018\,(0.018)$ \\
	K\,{\sc i} & 2 & $0.032\,(0.034)$ & 2 & $0.030\,(0.033)$ \\
	Ca\,{\sc i} & 14 & $0.014\,(0.014)$ & 16 & $0.014\,(0.015)$ \\
	Ca\,{\sc ii} & 5 & $0.030\,(0.027)$ & 2 & $0.066\,(0.064)$ \\
	Ti\,{\sc i} & 9 & $0.015\,(0.015)$ & 6 & $0.021\,(0.022)$ \\
	Cr\,{\sc i} & 10 & $0.026\,(0.030)$ & 9 & $0.028\,(0.032)$ \\
	Mn\,{\sc i} & 2 & $0.044\,(0.043)$ & 2 & $0.038\,(0.035)$ \\
	Fe\,{\sc i} & 103 & $0.003\,(0.004)$ & 89 & $0.004\,(0.004)$ \\
	Fe\,{\sc ii} & 5 & $0.027\,(0.030)$ & 4 & $0.040\,(0.045)$ \\
	Ni\,{\sc i} & 9 & $0.010\,(0.010)$ & 8 & $0.014\,(0.015)$ \\
	Zn\,{\sc i} & 1 & $0.117\,(0.116)$ & 1 & $0.109\,(0.107)$ \\
	Sr\,{\sc ii} & 2 & $0.039\,(0.041)$ & 2 & $0.035\,(0.038)$ \\
\hline
Total & 256 & ~ & 237 & ~ \\
\enddata
\tablecomments{Mean abundance errors indicate the mean half-widths of the uncertainty-like intervals for individual elements over calibrator stars, with the corresponding values for non-calibrators given in parentheses.}
\end{deluxetable}

\subsection{Elemental Abundances of Solar Analogs} \label{subsec:abundances}

For each combination of star, element, and line list (VALD or MB99), we combined
the accepted per-line residual curves after applying the per-line zero-point shifts
and scatter propagation described in Section~\ref{subsec:estimation}.
Unless stated otherwise, we adopt abundances from neutral lines (X\,{\sc i})
for each element, except for Sr, for which we use Sr\,{\sc ii}.

As an additional quality-control step, we compared each target spectrum with the
combined spectrum of the calibrators and rejected a line for a given star if
one or more pixels within $\pm 15$~km~s$^{-1}$ of the line deviated by more than
$3\times$ the local standard deviation from the combined spectrum (Section~\ref{subsec:combined_spec}).
We also flagged pixels with normalized flux greater than 1.05 in the line window,
since no emission feature is expected in solar analogs.
Most of these rejections were associated with residual telluric features near the edges
of the $Y$ and $J$ bands, and they were more frequent in spectra from the 2025 February run,
which was affected by stronger telluric absorption (Appendix~\ref{sec:telluric}).
For phosphorus, only a few cases of a small number of stars were rejected, and no star
lost more than one P\,{\sc i} line, so the impact on the final [P/H] estimates is minor.

Table~\ref{tab:abundances} presents a subset of the resulting abundance measurements,
while the full machine-readable table, including all elements and both line lists,
is provided online.
\textit{Two-sided} constraints (i.e., neither upper nor lower limits) were obtained for most elements
and stars. The main exception was Zn, for which the only available Zn\,{\sc i} line
at 13053.64\,{\AA} is weak in these stars and yielded only upper limits for 7 stars.
Table~\ref{tab:lines_stats} lists the mean values of errors $e_\mathrm{X}$
defined in Section~\ref{subsec:estimation}, $\langle e_\mathrm{X}\rangle$,
for both calibrators and non-calibrators.
There is no significant difference in $\langle e_\mathrm{X}\rangle$ between the two groups,
supporting the internal consistency of the abundance analysis across the full solar-analog sample.
Most species yield typical errors below $\sim$0.05~dex, whereas species represented by
fewer or weaker lines show broader intervals.

\begin{deluxetable*}{lccccc}
\tablecaption{Detailed Abundances of Solar Analogs\label{tab:abundances}}
\tablehead{\colhead{Star} & \colhead{[Fe/H]} & \colhead{[Mg/H]} & \colhead{[Si/H]} & \colhead{[P/H]} & \colhead{[S/H]}}
\startdata
HIP\,1954 & $-0.087 \pm 0.003\,(103)$ & $-0.057 \pm 0.012\,(8)$ & $-0.086 \pm 0.003\,(55)$ & $-0.077 \pm 0.022\,(5)$ & $-0.123 \pm 0.025\,(3)$ \\
HIP\,3203 & $-0.075 \pm 0.007\,(92)$ & $-0.131 \pm 0.026\,(8)$ & $-0.026 \pm 0.005\,(53)$ & $-0.223 \pm 0.072\,(5)$ & $-0.133 \pm 0.044\,(3)$ \\
HIP\,4909 & $+0.033 \pm 0.002\,(103)$ & $+0.023 \pm 0.011\,(8)$ & $+0.070 \pm 0.003\,(55)$ & $-0.151 \pm 0.014\,(5)$ & $+0.037 \pm 0.024\,(3)$ \\
HIP\,5301 & $-0.085 \pm 0.002\,(103)$ & $-0.053 \pm 0.011\,(8)$ & $-0.070 \pm 0.004\,(55)$ & $-0.093 \pm 0.035\,(5)$ & $-0.069 \pm 0.026\,(3)$ \\
HIP\,6407 & $-0.067 \pm 0.002\,(103)$ & $-0.127 \pm 0.011\,(8)$ & $-0.036 \pm 0.004\,(55)$ & $-0.215 \pm 0.033\,(5)$ & $-0.097 \pm 0.035\,(3)$ \\
HIP\,7585 & $+0.109 \pm 0.002\,(103)$ & $+0.059 \pm 0.014\,(8)$ & $+0.082 \pm 0.005\,(55)$ & $+0.079 \pm 0.021\,(5)$ & $+0.035 \pm 0.018\,(3)$ \\
HIP\,8507 & $-0.103 \pm 0.003\,(102)$ & $-0.077 \pm 0.010\,(8)$ & $-0.100 \pm 0.004\,(55)$ & $-0.109 \pm 0.031\,(5)$ & $-0.157 \pm 0.024\,(3)$ \\
HIP\,9349 & $-0.001 \pm 0.002\,(103)$ & $-0.087 \pm 0.014\,(8)$ & $+0.008 \pm 0.005\,(55)$ & $-0.103 \pm 0.029\,(5)$ & $+0.025 \pm 0.021\,(3)$ \\
HIP\,10175 & $-0.005 \pm 0.002\,(103)$ & $-0.049 \pm 0.011\,(8)$ & $+0.000 \pm 0.005\,(55)$ & $-0.027 \pm 0.020\,(5)$ & $+0.053 \pm 0.014\,(3)$ \\
HIP\,10303 & $+0.117 \pm 0.002\,(102)$ & $+0.135 \pm 0.007\,(7)$ & $+0.116 \pm 0.003\,(55)$ & $+0.075 \pm 0.034\,(5)$ & $+0.109 \pm 0.036\,(3)$ \\
HIP\,14501 & $-0.133 \pm 0.002\,(103)$ & $+0.015 \pm 0.007\,(8)$ & $-0.062 \pm 0.003\,(55)$ & $+0.007 \pm 0.025\,(5)$ & $-0.045 \pm 0.018\,(3)$ \\
HIP\,14614 & $-0.117 \pm 0.003\,(100)$ & $-0.077 \pm 0.007\,(8)$ & $-0.102 \pm 0.003\,(55)$ & $-0.121 \pm 0.032\,(5)$ & $-0.147 \pm 0.020\,(3)$ \\
HIP\,15527 & $-0.051 \pm 0.003\,(103)$ & $-0.057 \pm 0.013\,(8)$ & $-0.070 \pm 0.004\,(55)$ & $-0.081 \pm 0.024\,(5)$ & $-0.085 \pm 0.025\,(3)$ \\
HIP\,18844 & $-0.007 \pm 0.003\,(103)$ & $+0.051 \pm 0.012\,(8)$ & $+0.026 \pm 0.005\,(55)$ & $-0.055 \pm 0.029\,(5)$ & $+0.143 \pm 0.031\,(3)$ \\
HIP\,30037 & $-0.017 \pm 0.005\,(97)$ & $+0.045 \pm 0.015\,(8)$ & $-0.030 \pm 0.007\,(51)$ & $-0.057 \pm 0.024\,(5)$ & $-0.085 \pm 0.028\,(3)$ \\
HIP\,30158 & $+0.005 \pm 0.003\,(97)$ & $+0.073 \pm 0.012\,(8)$ & $+0.012 \pm 0.007\,(51)$ & $-0.067 \pm 0.054\,(5)$ & $+0.047 \pm 0.014\,(3)$ \\
HIP\,30502 & $-0.049 \pm 0.003\,(100)$ & $+0.015 \pm 0.012\,(8)$ & $-0.082 \pm 0.007\,(52)$ & $-0.069 \pm 0.030\,(5)$ & $-0.107 \pm 0.012\,(3)$ \\
HIP\,34511 & $-0.113 \pm 0.004\,(86)$ & $-0.145 \pm 0.014\,(8)$ & $-0.084 \pm 0.009\,(38)$ & $-0.145 \pm 0.056\,(4)$ & $-0.115 \pm 0.025\,(3)$ \\
HIP\,36515 & $-0.085 \pm 0.007\,(81)$ & $+0.081 \pm 0.008\,(7)$ & $-0.058 \pm 0.009\,(45)$ & $-0.095 \pm 0.035\,(4)$ & $-0.041 \pm 0.027\,(3)$ \\
HIP\,38072 & $+0.107 \pm 0.004\,(95)$ & $+0.037 \pm 0.014\,(8)$ & $+0.122 \pm 0.006\,(49)$ & $+0.017 \pm 0.054\,(4)$ & $+0.051 \pm 0.011\,(3)$ \\
HIP\,41317 & $-0.083 \pm 0.004\,(102)$ & $-0.037 \pm 0.016\,(7)$ & $-0.076 \pm 0.007\,(54)$ & $-0.083 \pm 0.042\,(5)$ & $-0.027 \pm 0.039\,(3)$ \\
HIP\,43297 & $+0.105 \pm 0.006\,(91)$ & $-0.055 \pm 0.023\,(7)$ & $+0.026 \pm 0.006\,(50)$ & $+0.015 \pm 0.038\,(5)$ & $+0.015 \pm 0.025\,(3)$ \\
HIP\,44713 & $+0.095 \pm 0.005\,(91)$ & $+0.023 \pm 0.023\,(8)$ & $+0.032 \pm 0.007\,(47)$ & $+0.155 \pm 0.041\,(4)$ & $+0.181 \pm 0.023\,(3)$ \\
HIP\,44935 & $+0.027 \pm 0.005\,(99)$ & $+0.105 \pm 0.015\,(8)$ & $+0.072 \pm 0.007\,(54)$ & $+0.039 \pm 0.037\,(5)$ & $-0.031 \pm 0.014\,(3)$ \\
HIP\,44997 & $+0.015 \pm 0.005\,(91)$ & $-0.073 \pm 0.007\,(8)$ & $-0.026 \pm 0.007\,(51)$ & $-0.019 \pm 0.051\,(5)$ & $-0.065 \pm 0.028\,(3)$ \\
HIP\,49756 & $-0.019 \pm 0.002\,(103)$ & $+0.007 \pm 0.009\,(8)$ & $+0.030 \pm 0.004\,(54)$ & $+0.087 \pm 0.035\,(4)$ & $-0.031 \pm 0.017\,(3)$ \\
HIP\,54102 & $+0.013 \pm 0.005\,(101)$ & $+0.015 \pm 0.008\,(8)$ & $+0.022 \pm 0.006\,(54)$ & $-0.159 \pm 0.037\,(5)$ & $+0.101 \pm 0.021\,(3)$ \\
HIP\,54287 & $+0.141 \pm 0.003\,(94)$ & $+0.183 \pm 0.007\,(6)$ & $+0.114 \pm 0.004\,(47)$ & $-0.021 \pm 0.038\,(5)$ & $+0.119 \pm 0.033\,(3)$ \\
HIP\,54582 & $-0.117 \pm 0.004\,(102)$ & $-0.037 \pm 0.010\,(8)$ & $-0.098 \pm 0.005\,(55)$ & $-0.107 \pm 0.023\,(5)$ & $-0.087 \pm 0.025\,(3)$ \\
HIP\,62039 & $+0.071 \pm 0.002\,(92)$ & $+0.159 \pm 0.017\,(7)$ & $+0.076 \pm 0.002\,(49)$ & $+0.191 \pm 0.033\,(5)$ & $+0.133 \pm 0.016\,(3)$ \\
HIP\,64673 & $-0.019 \pm 0.004\,(102)$ & $+0.057 \pm 0.009\,(8)$ & $+0.026 \pm 0.006\,(54)$ & $+0.023 \pm 0.027\,(5)$ & $+0.061 \pm 0.017\,(3)$ \\
HIP\,65708 & $-0.097 \pm 0.003\,(97)$ & $+0.033 \pm 0.018\,(7)$ & $+0.004 \pm 0.006\,(52)$ & $+0.009 \pm 0.036\,(5)$ & $+0.045 \pm 0.033\,(3)$ \\
HIP\,74389 & $+0.195 \pm 0.004\,(92)$ & $+0.027 \pm 0.017\,(8)$ & $+0.144 \pm 0.006\,(50)$ & $+0.077 \pm 0.026\,(5)$ & $+0.041 \pm 0.036\,(3)$ \\
HIP\,96160 & $-0.049 \pm 0.003\,(103)$ & $-0.085 \pm 0.005\,(8)$ & $-0.082 \pm 0.007\,(55)$ & $-0.101 \pm 0.038\,(5)$ & $-0.057 \pm 0.018\,(3)$ \\
HIP\,101905 & $+0.069 \pm 0.004\,(99)$ & $+0.037 \pm 0.013\,(8)$ & $+0.050 \pm 0.006\,(54)$ & $-0.153 \pm 0.014\,(5)$ & $+0.005 \pm 0.018\,(3)$ \\
HIP\,102152 & $-0.037 \pm 0.004\,(103)$ & $-0.047 \pm 0.009\,(8)$ & $-0.036 \pm 0.004\,(55)$ & $-0.011 \pm 0.022\,(5)$ & $+0.021 \pm 0.026\,(3)$ \\
HIP\,104045 & $+0.071 \pm 0.003\,(103)$ & $-0.027 \pm 0.010\,(8)$ & $+0.034 \pm 0.005\,(55)$ & $-0.029 \pm 0.043\,(5)$ & $+0.077 \pm 0.035\,(3)$ \\
HIP\,105184 & $-0.011 \pm 0.003\,(103)$ & $-0.019 \pm 0.011\,(8)$ & $-0.050 \pm 0.006\,(55)$ & $-0.253 \pm 0.029\,(5)$ & $-0.063 \pm 0.027\,(3)$ \\
HIP\,108158 & $+0.043 \pm 0.004\,(97)$ & $+0.195 \pm 0.013\,(5)$ & $+0.132 \pm 0.006\,(51)$ & $+0.101 \pm 0.054\,(5)$ & $+0.089 \pm 0.030\,(3)$ \\
HIP\,109821 & $-0.119 \pm 0.003\,(103)$ & $-0.023 \pm 0.010\,(8)$ & $+0.010 \pm 0.003\,(55)$ & $-0.037 \pm 0.016\,(5)$ & $-0.065 \pm 0.018\,(3)$ \\
HIP\,114328 & $-0.023 \pm 0.004\,(103)$ & $+0.001 \pm 0.010\,(8)$ & $-0.022 \pm 0.005\,(55)$ & $+0.031 \pm 0.022\,(5)$ & $-0.057 \pm 0.025\,(3)$ \\
HIP\,114615 & $-0.139 \pm 0.004\,(103)$ & $-0.101 \pm 0.009\,(8)$ & $-0.100 \pm 0.007\,(55)$ & $-0.087 \pm 0.034\,(5)$ & $-0.095 \pm 0.021\,(3)$ \\
HIP\,115577 & $+0.027 \pm 0.003\,(101)$ & $+0.175 \pm 0.015\,(6)$ & $+0.150 \pm 0.006\,(53)$ & $+0.077 \pm 0.033\,(5)$ & $+0.039 \pm 0.021\,(3)$ \\
HIP\,116906 & $-0.007 \pm 0.001\,(103)$ & $+0.043 \pm 0.008\,(8)$ & $+0.016 \pm 0.004\,(55)$ & $-0.019 \pm 0.018\,(5)$ & $+0.041 \pm 0.020\,(3)$ \\
HIP\,117367 & $+0.021 \pm 0.001\,(103)$ & $+0.069 \pm 0.008\,(8)$ & $+0.052 \pm 0.004\,(55)$ & $+0.089 \pm 0.018\,(5)$ & $+0.069 \pm 0.027\,(3)$ \\
HIP\,118115 & $-0.035 \pm 0.003\,(103)$ & $+0.063 \pm 0.013\,(8)$ & $+0.016 \pm 0.003\,(55)$ & $-0.035 \pm 0.019\,(5)$ & $+0.049 \pm 0.031\,(3)$ \\
\enddata
\tablecomments{Values measured with the VALD line list after the per-line calibration are presented here for five elements.
Parentheses give the number of residual curves used. The full machine-readable table is available online.}
\end{deluxetable*}

\subsubsection{Validation of Derived Abundances and Uncertainties} \label{subsubsec:abundance_validation}

\begin{figure}[!tbp]
\centering
\includegraphics[width=0.90\linewidth]{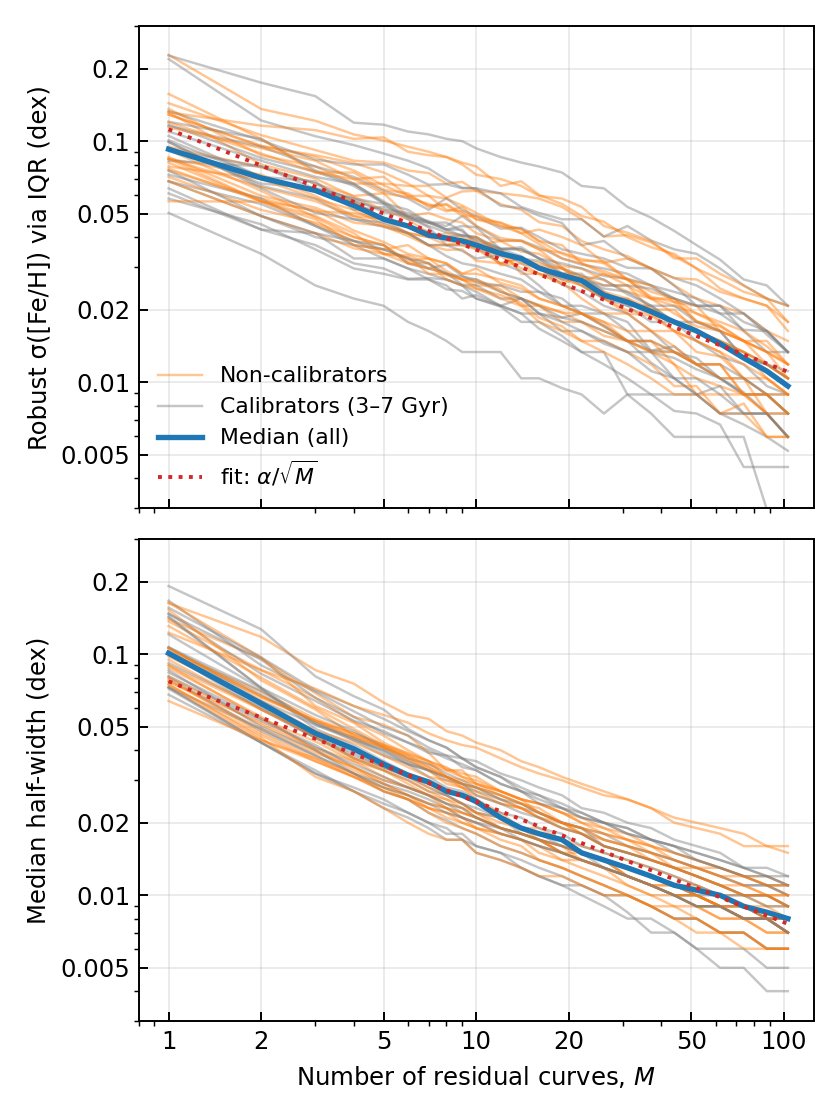}
\caption{Bootstrap test using Fe\,{\sc i} VALD combined curves for solar-analog stars.
Gray and orange lines show,
respectively for individual calibrators (ages 3--7~Gyr) and non-calibrators,
the robust $\sigma$([Fe/H]) and median half-width versus subset size $M$;
blue curves show the median over all stars;
red curves are $\alpha/\sqrt{M}$ fits. Both metrics decline approximately as
expected from statistical averaging.}
\label{fig:feh_combined_sim}
\end{figure}

To test whether the uncertainty estimates from the combined residual curves provide a
statistically meaningful precision scale, we performed a bootstrap experiment using
$\sim$100 Fe\,{\sc i} residual curves for each solar-analog star.
For each star and each subset size $M$ ($\leq 100$), we performed 1000 random draws
of $M$ residual curves from the available Fe\,{\sc i} lines to
derive trial [Fe/H] values, and measured
(i)~the median half-width of the resulting intervals and
(ii)~the scatter of the recovered [Fe/H] values using the interquartile range
($\sigma = 0.74 \times \mathrm{IQR}$).
If the combined-curve intervals behave statistically, both quantities should decrease
approximately as $\alpha/\sqrt{M}$ with increasing $M$, where $\alpha$ is a
star-dependent scaling factor. Figure~\ref{fig:feh_combined_sim} shows this expected behavior,
indicating that the combined-residual intervals provide reasonable
statistical uncertainty estimates.
For very large $M$, however, the formal intervals may become unrealistically small,
and the quoted uncertainties should therefore be interpreted as internal statistical errors
rather than complete error budgets.

\begin{figure}[!tbp]
\centering
\includegraphics[width=0.9\linewidth]{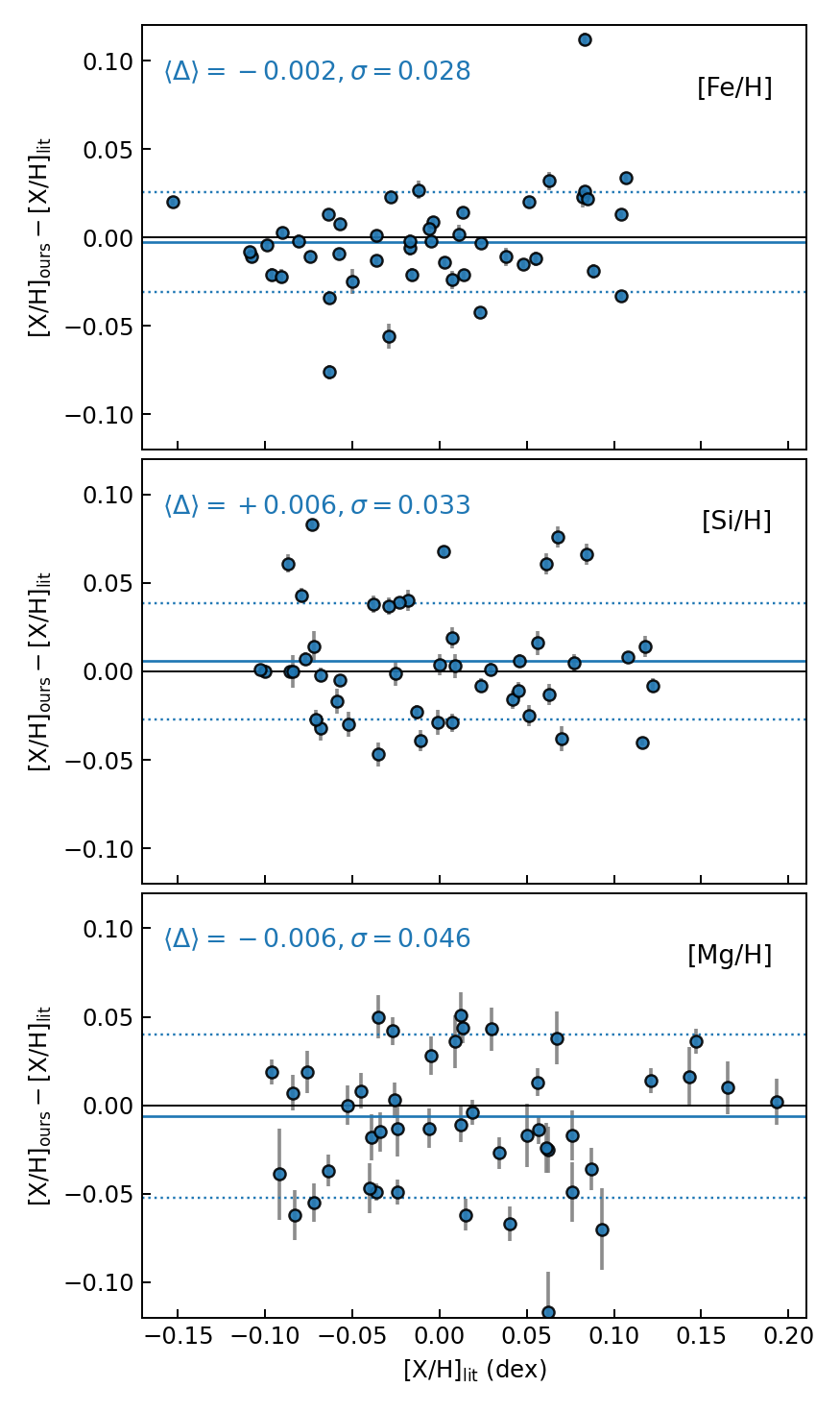}
\caption{Residuals between our VALD abundances and literature values for Fe\,{\sc i},
Si\,{\sc i}, and Mg\,{\sc i}, plotted versus literature [X/H]. Error bars indicate our
1$\sigma$ intervals, and the horizontal line marks zero residual.}
\label{fig:abund_vs_lit_in_paper}
\end{figure}

Our abundance measurements also agree well with the literature values from
\citet{Bedell-2018}.
Table~\ref{tab:abund_lit_delta} summarizes the mean residuals $\langle\Delta\rangle$
and scatters $\sigma(\Delta)$ between our abundances and the literature values for each element.
For most species, the residual scatter is well below $\sim$0.1~dex, whereas elements
represented by only a few lines (e.g., Na, Al, and Zn) show larger scatters.
Figure~\ref{fig:abund_vs_lit_in_paper} compares our VALD abundances with the optical
results of \citet{Bedell-2018} for Fe\,{\sc i}, Si\,{\sc i}, and Mg\,{\sc i}, showing
good agreement with scatters of 0.03--0.05~dex.
Such residuals are comparable to those found among high-quality optical studies themselves,
for example between \citet{Spina-2018} and \citet{Ramirez-2014a}.

\begin{deluxetable*}{lcccccc}
\tablecaption{Residuals relative to literature abundances\label{tab:abund_lit_delta}}
\tablehead{\colhead{Element} & \multicolumn{3}{c}{VALD} & \multicolumn{3}{c}{MB99} \\ 
 & \colhead{$\langle\Delta\rangle$} & \colhead{$\sigma(\Delta)$} & \colhead{$N_\mathrm{star}$} & \colhead{$\langle\Delta\rangle$} & \colhead{$\sigma(\Delta)$} & \colhead{$N_\mathrm{star}$}}
\startdata
C  & $-0.013$ & $0.063$ & 46 & $+0.013$ & $0.050$ & 46 \\
Na & $+0.029$ & $0.159$ & 44 & $+0.015$ & $0.082$ & 46 \\
Mg & $-0.006$ & $0.046$ & 46 & $-0.002$ & $0.049$ & 46 \\
Al & $+0.002$ & $0.133$ & 44 & $-0.005$ & $0.141$ & 44 \\
Si & $+0.006$ & $0.033$ & 46 & $-0.001$ & $0.031$ & 46 \\
S  & $+0.001$ & $0.056$ & 46 & $-0.005$ & $0.052$ & 46 \\
Ca & $-0.011$ & $0.067$ & 46 & $+0.000$ & $0.056$ & 46 \\
Ti & $+0.002$ & $0.049$ & 46 & $-0.005$ & $0.051$ & 46 \\
Cr & $-0.025$ & $0.085$ & 46 & $-0.002$ & $0.092$ & 46 \\
Mn & $-0.043$ & $0.102$ & 44 & $-0.035$ & $0.094$ & 44 \\
Fe & $-0.002$ & $0.028$ & 46 & $+0.004$ & $0.032$ & 46 \\
Ni & $+0.009$ & $0.031$ & 46 & $+0.007$ & $0.055$ & 46 \\
Zn & $-0.040$ & $0.187$ & 36 & $-0.065$ & $0.181$ & 36 \\
Sr & $-0.005$ & $0.053$ & 46 & $+0.001$ & $0.050$ & 46 \\
\enddata
\tablecomments{Means and scatters of $[\mathrm{{X}}/\mathrm{{H}}]_{\mathrm{{ours}}} - [\mathrm{{X}}/\mathrm{{H}}]_{\mathrm{{lit}}}$, shown separately for the analysis based on VALD and MB99. The literature abundances are taken from \citet{Bedell-2018} and \citet{Spina-2018}.}
\end{deluxetable*}

\subsubsection{Constraints on Phosphorus Abundances} \label{subsubsec:p_constraints}

\begin{figure}[!tbp]
\centering
\includegraphics[width=0.95\linewidth]{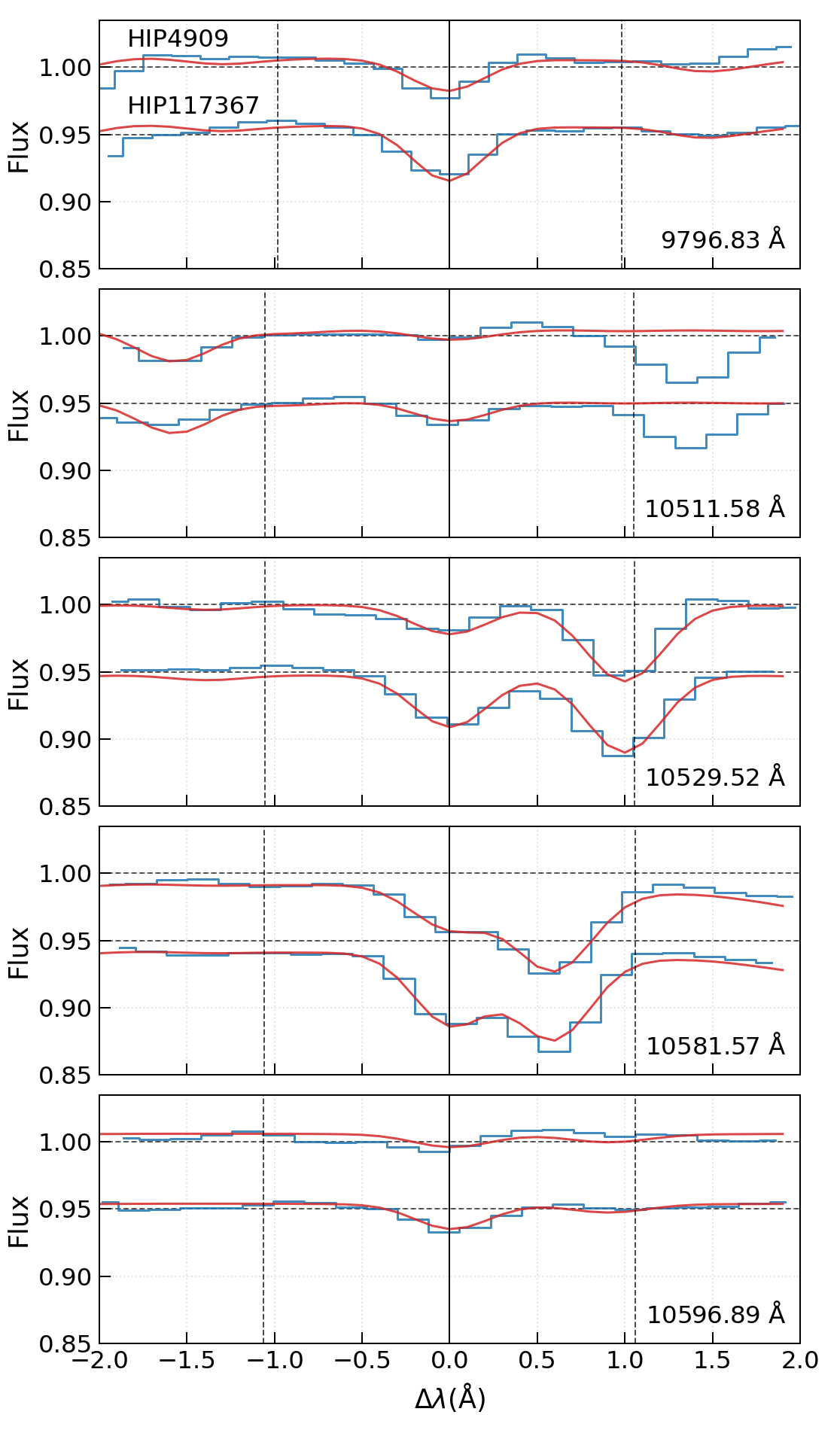}
\caption{Five P\,{\sc i} lines we selected. Observed spectra are shown in blue and
the best-fitting synthetic spectra are shown in red for two solar analogs,
HIP\,4909 (upper) and HIP\,117367 (lower). The solid vertical line marks the
line center, and the dashed vertical lines indicate the fitting range (corresponding to $\pm 30$~km~s$^{-1}$).}
\label{fig:p1_lines_example}
\end{figure}

As phosphorus is a key target of this study, we summarize the constraints
from individual P\,{\sc i} lines in Table~\ref{tab:p1_constraints}.
Five P\,{\sc i} lines from VALD and four from MB99 pass our selection criteria;
the five lines are illustrated in Figure~\ref{fig:p1_lines_example} for two solar analogs,
with relatively low (HIP\,4909, [P/H]$=-0.15$)
or high (HIP\,117367, [P/H]$=+0.07$) phosphorus abundances; the figure shows the observed spectra and the best-fitting synthetic spectra for each line.
The line at 9796.83\,{\AA}\ is included only in VALD simply because of the limited
wavelength coverage of MB99, while the remaining lines are common to both line lists.
Among these lines, P\,{\sc i} 10581.57\,{\AA}\ provides
the strongest constraints for all solar analogs in our sample, yielding abundance errors of
$\sim 0.07$~dex on average.
The remaining P\,{\sc i} lines also provide reasonable constraints,
with median half-widths of 0.07--0.17~dex.
As a result, the combined P\,{\sc i} abundances achieve median half-widths of 0.03~dex (VALD) and
0.04~dex (MB99) across the solar analogs.
The typical absorption of the strongest P\,{\sc i} 10581.57\,{\AA}\ line is only $\sim 0.05$
in depth ($\sim$30~m{\AA}\  in equivalent width),
underscoring the intrinsic difficulty of measuring phosphorus from weak lines in these stars.
Nevertheless, the combined results demonstrate that precise phosphorus abundances
can be obtained through the use of multiple lines and careful empirical calibration.

\begin{deluxetable}{lrrrr}
\tablecaption{Median Half-Width Constraints for P\,{\sc i}\label{tab:p1_constraints}}
\tablehead{Line & \multicolumn{2}{c}{VALD} & \multicolumn{2}{c}{MB99} \\
  \cmidrule(r){2-3} \cmidrule(l){4-5}
  & \colhead{$\sigma_\mathrm{median}$} & \colhead{$N_\mathrm{star}$} & \colhead{$\sigma_\mathrm{median}$} & \colhead{$N_\mathrm{star}$} }
\startdata
P\,{\sc i} 9796.83 & 0.083 & 38 & \multicolumn{1}{c}{---} & 0 \\
P\,{\sc i} 10511.58 & 0.165 & 38 & 0.152 & 37 \\
P\,{\sc i} 10529.52 & 0.073 & 44 & 0.114 & 44 \\
P\,{\sc i} 10581.57 & 0.069 & 46 & 0.065 & 46 \\
P\,{\sc i} 10596.89 & 0.123 & 43 & 0.117 & 42 \\
\hline
P\,{\sc i} combined & 0.033 & 46 & 0.039 & 46 \\
\enddata
\tablecomments{The half-width $\sigma_\mathrm{median}$ measures $0.5\times$(high -- low).
$N_\mathrm{star}$ is the number of stars with residual curves yielding
\textit{two-sided} constraints with full widths narrower than the width limit,
0.75~dex, used for calculating the median here.}
\end{deluxetable}

\section{Discussion} \label{sec:discussion}

\subsection{Age--[X/Fe] Relations} \label{subsec:age_relations}

We investigate age--[X/Fe] relations for all elements, other than Fe\,{\sc i}, measured in this study
by fitting linear trends to the solar-analog sample.
When measurements are available for both neutral and ionized species,
we adopt the results from the neutral species
(i.e., Fe\,{\sc i} and Ca\,{\sc i} rather than Fe\,{\sc ii} and Ca\,{\sc ii}),
while [Sr/Fe] is from Sr\,{\sc ii} in our measurements. 
For [X/Fe], we use [Fe/H] from the literature \citep{Bedell-2018}
rather than our [Fe/H] measurements, which still agree well with literature values
(Figure~\ref{fig:abund_vs_lit_in_paper}).

\begin{figure*}[!tbp]
\centering
\includegraphics[width=0.78\linewidth]{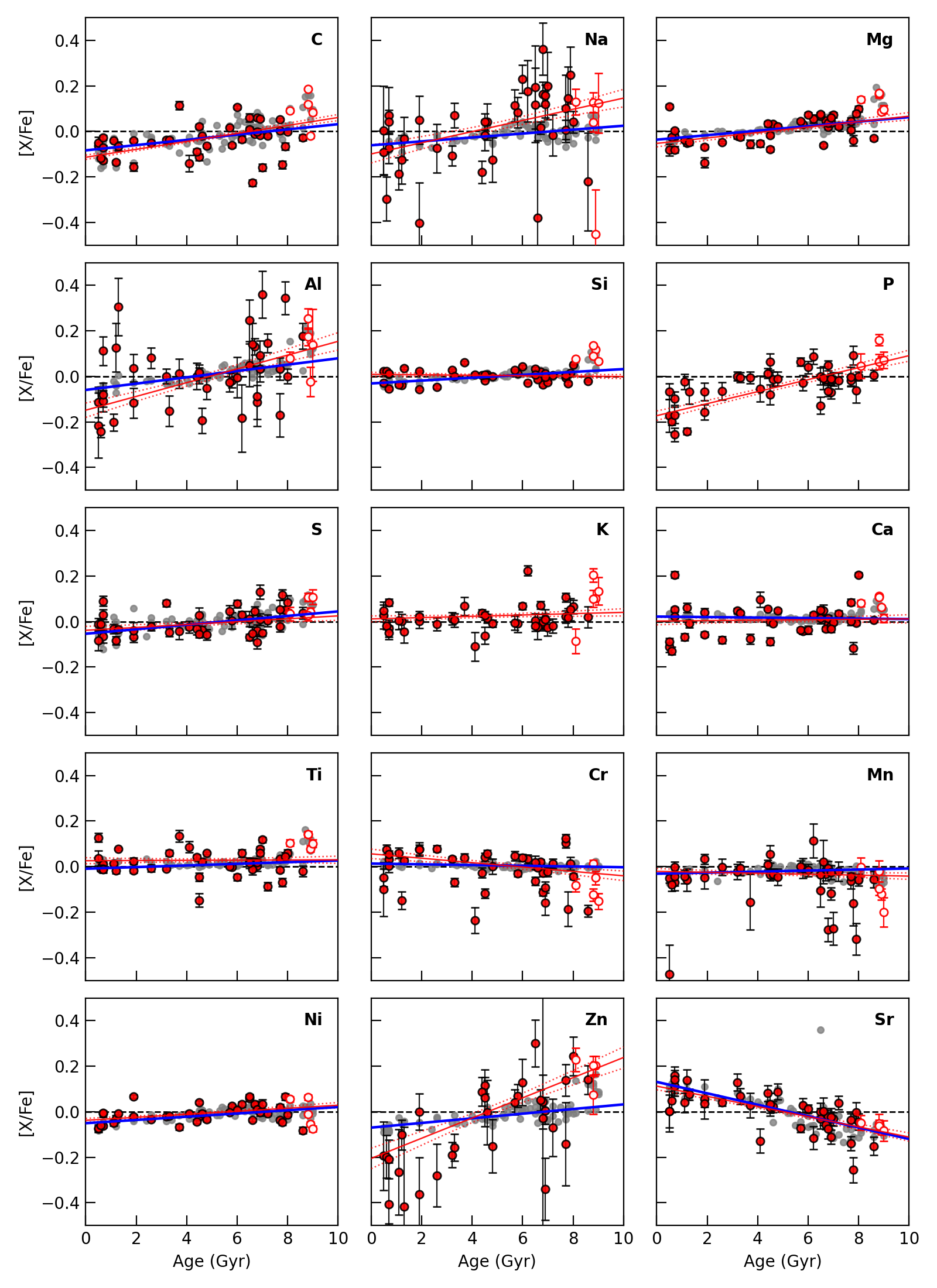}
\caption{Age--[X/Fe] relations for 15 elements based on the analysis using VALD after the calibration.
[X/Fe] obtained for neutral species and Sr\,{\sc ii} are shown with measurement uncertainties; 
red and blue solid lines mark our fits and those by \citet{Bedell-2018}, 
respectively. White circles indicate stars not used in the fitting (see text).}
\label{fig:age_xfe_panels}
\end{figure*}

\begin{figure}[!tbp]
\centering
\includegraphics[width=0.93\linewidth]{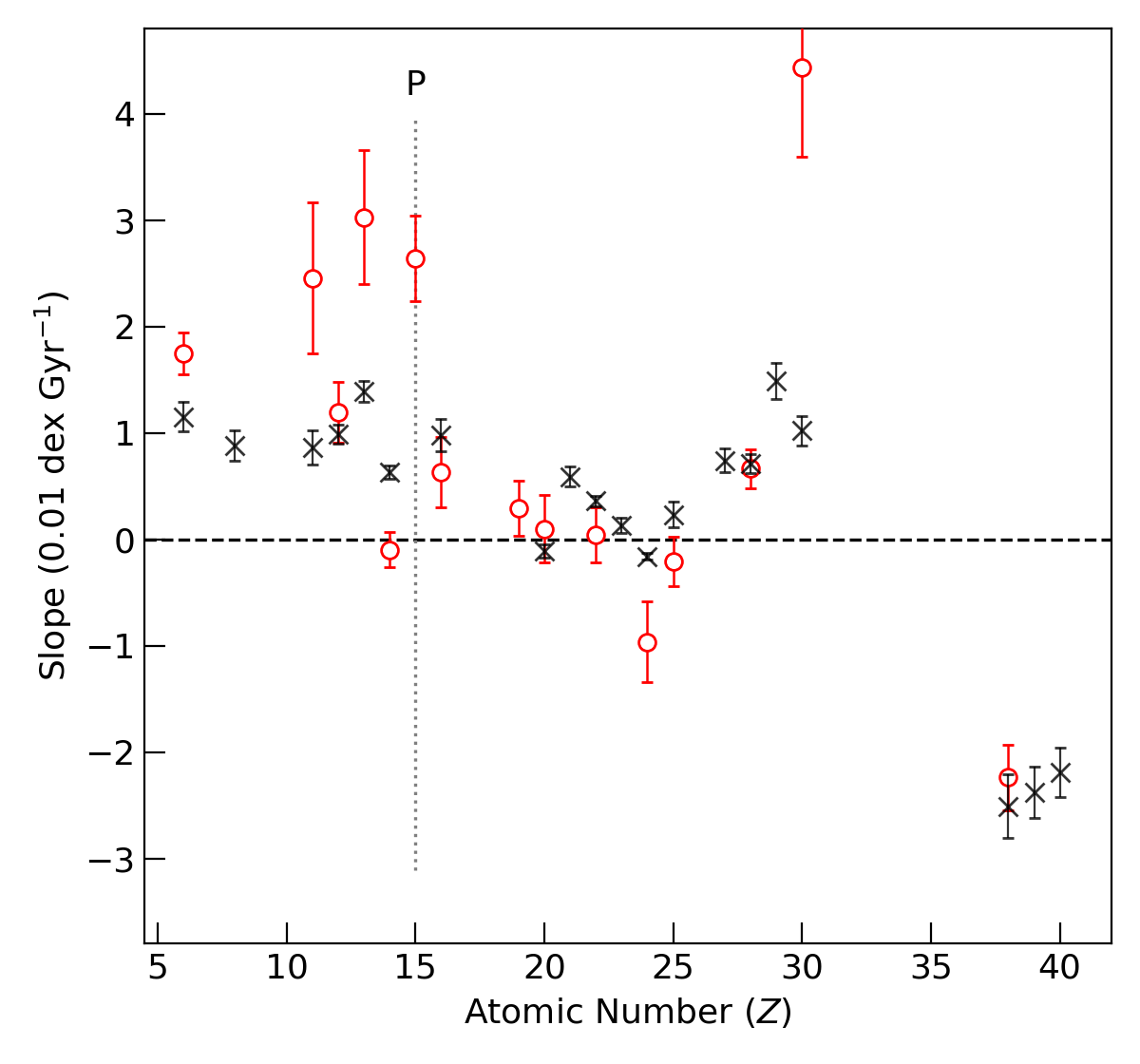}
\caption{Comparison of our age--[X/Fe] slopes (red) 
against the literature slopes (black) from \citet{Bedell-2018}.}
\label{fig:compare_slopes_vald}
\end{figure}

Following \citet{Bedell-2018}, we exclude stars belonging to 
different populations from the fitting of the age--[X/Fe] relations;
five stars (HIP\,14501, HIP\,65708, HIP\,108158, HIP\,109821, and HIP\,115577),
all with age higher than 8~Gyr, were excluded and indicated by white circles
in Figure~\ref{fig:age_xfe_panels}.
We made weighted least-squares fits to the data,
taking into account the uncertainties in [X/Fe],
with 2$\sigma$ clipping to remove outliers once.
Up to four measurements were rejected for a given element in this process.
The results based on VALD are shown in Figure~\ref{fig:age_xfe_panels}, while 
those obtained using MB99 are very similar. 
The best-fit slopes, in the unit of 0.01~dex~Gyr$^{-1}$, are summarized
in Table~\ref{tab:age_slopes}, and 
compared with literature values from \citet{Bedell-2018}
in Figure~\ref{fig:compare_slopes_vald}.
For Sr, note that \citet{Bedell-2018} adopted the measurements
by \citet{Spina-2018} in which Sr\,{\sc i} lines were used,
whereas our trends are based on Sr\,{\sc ii} lines.

Where a direct comparison is possible,  
our slopes are in reasonable agreement with those from \citet{Bedell-2018},
typically within 2~$\sigma$ uncertainties. 
The inferred age--[X/Fe] slopes for Na, Al, and Zn are, however, poorly constrained
with large uncertainties, so these trends should be treated as tentative.
In contrast, the slope of the age--[Si/Fe] relation appears to be significantly
different between our work and \citet{Bedell-2018}; our result indicates
a nearly flat trend, while \citet{Bedell-2018} found a moderately positive slope.
The origin of this discrepancy is unclear, but it may be related to
differences in the line lists and the systematics between optical and near-infrared analyses.

Our work successfully builds upon previous optical studies
by obtaining age--[X/Fe] relations for P and K.
At the same time, however, the near-infrared wavelength coverage of WINERED limits access
to some elements typically studied in the optical. 
Although absorption lines of certain neutron-capture elements are present
in the $YJ$ bands \citep{Matsunaga-2020}, a large fraction of those lines are either too weak
or absent in solar analogs to provide useful abundance constraints.

\begin{deluxetable*}{lrrr}
\tablecaption{Slopes (0.01 dex~Gyr$^{-1}$) of Age--[X/Fe] Relations\label{tab:age_slopes}}
\tablehead{\colhead{Element} & \colhead{VALD} & \colhead{MB99} & \colhead{Bedell+18}}
\startdata
C  & $+1.75 \pm 0.20~(\sigma=0.03,~N=37)$ & $+1.89 \pm 0.23~(\sigma=0.04,~N=37)$ & $+1.15 \pm 0.14$ \\
Na & $+2.45 \pm 0.71~(\sigma=0.09,~N=36)$ & $+1.23 \pm 0.36~(\sigma=0.05,~N=37)$ & $+0.86 \pm 0.16$ \\
Mg & $+1.19 \pm 0.29~(\sigma=0.04,~N=39)$ & $+1.45 \pm 0.20~(\sigma=0.03,~N=39)$ & $+0.99 \pm 0.09$ \\
Al & $+3.03 \pm 0.63~(\sigma=0.09,~N=37)$ & $+3.31 \pm 0.68~(\sigma=0.10,~N=36)$ & $+1.39 \pm 0.10$ \\
Si & $-0.10 \pm 0.16~(\sigma=0.02,~N=38)$ & $-0.05 \pm 0.14~(\sigma=0.02,~N=39)$ & $+0.63 \pm 0.06$ \\
P  & $+2.64 \pm 0.40~(\sigma=0.07,~N=41)$ & $+1.90 \pm 0.41~(\sigma=0.06,~N=39)$ & \multicolumn{1}{c}{---} \\
S  & $+0.63 \pm 0.33~(\sigma=0.05,~N=39)$ & $+0.36 \pm 0.25~(\sigma=0.04,~N=39)$ & $+0.98 \pm 0.15$ \\
K  & $+0.29 \pm 0.26~(\sigma=0.04,~N=39)$ & $-0.02 \pm 0.21~(\sigma=0.03,~N=39)$ & \multicolumn{1}{c}{---} \\
Ca & $+0.10 \pm 0.32~(\sigma=0.04,~N=39)$ & $-0.25 \pm 0.24~(\sigma=0.03,~N=37)$ & $-0.11 \pm 0.06$ \\
Ti & $+0.04 \pm 0.26~(\sigma=0.04,~N=39)$ & $+0.32 \pm 0.29~(\sigma=0.05,~N=40)$ & $+0.36 \pm 0.05$ \\
Cr & $-0.96 \pm 0.38~(\sigma=0.05,~N=39)$ & $-0.10 \pm 0.41~(\sigma=0.06,~N=39)$ & $-0.16 \pm 0.03$ \\
Mn & $-0.21 \pm 0.23~(\sigma=0.04,~N=35)$ & $-0.24 \pm 0.23~(\sigma=0.04,~N=36)$ & $+0.23 \pm 0.12$ \\
Ni & $+0.66 \pm 0.18~(\sigma=0.03,~N=39)$ & $+0.48 \pm 0.24~(\sigma=0.03,~N=39)$ & $+0.71 \pm 0.09$ \\
Zn & $+4.44 \pm 0.84~(\sigma=0.10,~N=30)$ & $+4.23 \pm 0.65~(\sigma=0.08,~N=30)$ & $+1.02 \pm 0.14$ \\
Sr\tablenotemark{a} & $-2.24 \pm 0.31~(\sigma=0.04,~N=39)$ & $-2.25 \pm 0.30~(\sigma=0.04,~N=40)$ & $-2.51 \pm 0.30$ \\
\enddata
\tablenotetext{a}{Our results for Sr are based on the measurements with Sr\,{\sc ii} lines, while the value from \citet{Bedell-2018} is based on Sr\,{\sc i} measurements. For all the other elements, we consider measurements based on neutral atoms.}
\tablecomments{$\sigma$ is the weighted RMS scatter of the residuals about the linear fit in [X/Fe] (dex), and $N$ is the number of stars used in the fit after a 2\,$\sigma$ clipping.}
\end{deluxetable*}

Beyond the age--[X/Fe] relations, it is of interest to mention
the high sensitivity of [Sr/Mg] to age. These two elements show
opposite age trends, leading to the large slope of
$\mathrm{d[Sr/Mg]/dAge} \simeq -0.03\ \mathrm{dex~Gyr}^{-1}$ in our measurements.
In fact, many previous studies demonstrated that ratios of s-process elements
to $\alpha$ elements, especially [Y/Mg], are sensitive age indicators
\citep[e.g.,][]{Nissen-2015,Spina-2018,DelgadoMena-2019,Nissen-2020}.
For example, \citet{Nissen-2020} obtained a slope of
$-0.0346 \pm 0.0010~\mathrm{dex~Gyr}^{-1}$ for [Sr/Mg] in nearby solar-type stars,
which is in excellent agreement with our result.
\citet{Nissen-2020} also found that the residuals of the age--[Sr/Mg] relation
depend significantly on [Fe/H], whereas the corresponding [Y/Mg] residuals show
no significant metallicity dependence over $-0.3 < \mathrm{[Fe/H]} < +0.3$.
Readers are referred to \citet{Ratcliffe-2024,Shejeelammal-2024}
for recent discussions on the correlations between [s/$\alpha$] and age.

\subsection{Implications for Phosphorus in Galactic Chemical Evolution} \label{subsec:chemical_evolution}

Phosphorus shows one of the steepest age--abundance trends in our sample,
with $\mathrm{d[P/Fe]/dAge} \simeq +0.02\ \mathrm{dex~Gyr}^{-1}$,
significantly larger than those of the $\alpha$ elements, for which
the steepest case (Mg) reaches $\sim +0.01\ \mathrm{dex~Gyr}^{-1}$. 
The mild age dependence of [$\alpha$/Fe] represented by Mg is well understood in the framework
of Galactic chemical evolution (GCE), reflecting the balance between
prompt enrichment from CCSNe and delayed Fe production
from SNe~Ia controlled by the star-formation
timescale in the Galactic disk \citep[e.g.,][]{Matteucci-2001}.

The steeper age trend suggests that phosphorus does not behave like
a simple CCSN-dominated $\alpha$ element relative to Fe,
and likely reflects additional time-dependent contributions
and/or yield dependencies beyond the canonical CCSN+SNe Ia balance.
A good example is the s-process element Sr, which shows a steep negative slope,
$\mathrm{d[Sr/Fe]/dAge} \lesssim -0.02\ \mathrm{dex~Gyr}^{-1}$,
consistent with the dominant contribution of the main s-process
in low-mass AGB stars \citep{Spina-2018,Tsujimoto-2021}.
The opposite signs of the Sr and P slopes are consistent with different dominant
nucleosynthetic contributors.

For P, the steep positive slope does not by itself identify the dominant production site,
but it provides a robust empirical constraint for GCE models.
The amount of P synthesized before the disk reaches solar metallicity is expected to
depend on chemical-evolution history at each Galactic location,
leaving measurable differences even at [Fe/H] $\simeq 0$.
\citet{Maas-2022} suggested population-dependent [P/Fe]--[Fe/H] tracks,
and our pronounced age--[P/Fe] trend provides complementary evidence that
such differences do not vanish at solar metallicity.
Combining the age--[P/Fe] slope with
(i) low [P/Fe] in very metal-poor stars \citep{Roederer-2014} and
(ii) the non-monotonic [P/Fe]--[Fe/H] behavior (peaking at [P/Fe]$\simeq +0.5$ around [Fe/H]$=-1$~dex),
the observed evolution is difficult to reproduce with CCSNe and SNe Ia alone \citep[e.g.,][]{Bekki-2024}.
In the context of the candidate production channels summarized in the Introduction,
our steep age--[P/Fe] slope provides a stringent, age-resolved constraint for GCE models.

\section{Summary} \label{sec:summary}

We used WINERED WIDE-mode spectra of solar analogs to derive near-infrared abundances
for 16 elements, including Fe, and validated our measurements through line-by-line
calibration, internal consistency checks, and comparisons with the optical results
of \citet{Bedell-2018}. The abundance measurements were performed independently
for lines selected from two commonly-used atomic line lists, VALD and MB99.
After the empirical calibration, we retained 256 VALD lines and 237 MB99
lines for the abundance analysis, enabling robust measurements in the $YJ$ bands.

The resulting abundances show a level of precision comparable to that achieved in
optical high-resolution studies of solar analogs.
For elements with many usable lines, particularly Fe and Si, the internal uncertainties
derived from the combined residual curves are smaller than 0.01~dex.
Comparisons with literature values from optical spectroscopy demonstrate agreement
at the level of {$\sim$}0.02~dex, indicating that our near-infrared abundance scale is
consistent with established optical results.
As for phosphorus, which is a key element of this study, we detected five P\,{\sc i} lines 
and obtained precise abundances, with half-widths of the uncertainty intervals as small as $\sim$0.04~dex.
These findings demonstrate that high-resolution near-infrared spectroscopy can
deliver abundance measurements of solar analogs with a precision and accuracy
competitive with traditional optical analyses, while providing access to key
elements such as phosphorus that are difficult to measure at shorter wavelengths.

Linear fits to age--[X/Fe] relations broadly agree with literature slopes and extend
previous work by adding near-infrared constraints for 15 elements, including P and K,
with respect to Fe.
We found that phosphorus shows one of the steepest age--abundance trends
in our sample, with $\mathrm{d[P/Fe]/dAge} \simeq +0.02\ \mathrm{dex~Gyr}^{-1}$,
which is larger than those of the $\alpha$ elements.
In contrast, K, together with S, shows the flat trend, whose slope is 
barely significant with the error $\sim$ 0.003~dex~Gyr$^{-1}$.
Our results do not aim to identify the dominant P production site,
but they provide meaningful empirical constraints on GCE models.
Any successful model must reproduce the large age--[P/Fe] slope, its contrast with
the s-process behavior traced by Sr, and the non-monotonic evolution of [P/Fe]
with metallicity over a wide [Fe/H] range.

\begin{acknowledgments}
This work was supported by NINS Astrobiology Center Program Research
(Grant No.~AB0717) and the Toray Science and Technology Research Grant
from the Toray Science Foundation.
It is a great pleasure to thank the staff of Las Campanas Observatory (LCO)
for their support during the installation of WINERED and the observations
with the 6.5-m Magellan Clay Telescope at LCO, Chile.
We also thank the staff of Carnegie Observatories,
especially Andy McWilliam and Shreyas Vissapragada, for enabling
the operation of WINERED at LCO.
WINERED was developed by the University of Tokyo and the Laboratory of
Infrared High-resolution spectroscopy (LiH), Kyoto Sangyo University,
with financial support from Grants-in-Aid for Scientific Research
(KAKENHI) from the Japan Society for the Promotion of Science
(JSPS; Nos. 16684001, 20340042, and 21840052), and from the MEXT Supported
Program for the Strategic Research Foundation at Private Universities
(Nos.~S0801061 and S1411028).
The 2025 February observing run was partly supported by KAKENHI
(Grant No.~18H01248).
We are grateful to Valentina Ort\'uzar-Garz\'on for her assistance during the observations in 2025 August.
HS acknowledges support from JSPS KAKENHI Grant No.~19K03917.
SE acknowledges support from CONICYT/ANID FONDECYT Postdoctoral Fellow No.~324052.
MC is thankful for the support of the Universidad de Santiago de Chile DICYT project 042431PA Postdoc.

NM acknowledges the use of generative artificial intelligence tools during the
preparation of this work.
Specifically, ChatGPT was used to improve
the grammar, clarity, and readability of the manuscript, and also for
the code development and refinement (Codex).
No other generative AI tools were used.
The authors reviewed and edited all contents and take full responsibility
for the results and conclusions presented in this paper.
\end{acknowledgments}

\begin{contribution}

NM led the observations of solar analogs, data analysis, and manuscript preparation.
TT contributed to the discussion of chemical evolution.
DT and HS contributed to the discussion of the PCA-based telluric correction,
which was eventually implemented by NM.
SO, TT, and YS contributed to the operation of WINERED during the relevant observing runs.
DT, HS, IP, SE, and MC also contributed to the observations.
All authors contributed to the writing and revision of the manuscript.

\end{contribution}

%
\facilities{Magellan: Clay 6.5m (WINERED)}

\software{astropy \citep{astropy-2013,astropy-2018,astropy-2022},
          OCTOMAN \citep{Taniguchi-2025},
          WARP (\url{https://github.com/SatoshiHamano/WARP}),
          TerraPCA (\url{https://github.com/nmatsuna/winered-telluric-pca})
          }


\appendix

\section{PCA-Based Telluric Correction} \label{sec:telluric}

Telluric absorption in the WINERED WIDE-mode spectra was corrected using
the empirical PCA-based method implemented in TerraPCA.
Rather than relying on theoretical atmospheric transmission models,
TerraPCA constructs the telluric transmission spectrum directly from observed
spectra of hot, feature-poor standard stars.
Applications of PCA to telluric correction in high-resolution infrared spectroscopy
have been demonstrated previously \citep[e.g.,][]{Artigau-2014}.
The public implementation of TerraPCA is available at
\url{https://github.com/nmatsuna/winered-telluric-pca}.

For each spectral order, TerraPCA constructs an empirical telluric model from
standard-star spectra obtained over a range of airmass and atmospheric conditions.
We used $\sim$25 spectra from the 2017--2018 NTT runs and $\sim$15 spectra from
the 2022--2023 Magellan runs.
The spectra are interpolated onto a common wavelength grid, aligned in wavelength,
and decomposed with PCA.
The telluric transmission spectrum $T(\lambda)$ is then represented as a linear
combination of the resulting basis vectors.
Tests on standard-star spectra from 2024 and 2025 confirmed that these bases remain
effective for modeling telluric absorption at the $\sim$1\% level under a variety
of conditions,
which is adequate for the abundance analysis in the main text.

For each science spectrum, TerraPCA fits the telluric transmission as a linear
combination of the PCA bases plus a constant continuum level,
using least-squares optimization while masking pixels affected by strong stellar lines
or other unreliable features.
In the solar-analog analysis presented here, we first removed stellar absorption
approximately by dividing by the combined solar-analog spectrum, and then fitted the
telluric model to the resulting spectra to improve robustness (Section~\ref{sec:obs}).
An optional wavelength adjustment was applied in telluric-dominated regions when needed.
Telluric correction was not applied to orders m53 and m54 (10295--10685\,{\AA}),
where atmospheric absorption is negligible.

The quality of the correction was evaluated from the RMS of the residuals in normalized
flux over telluric-dominated wavelength regions.
This quantity measures how well the pixel-by-pixel structure of the telluric lines is
reproduced, although it can be slightly overestimated when stellar and telluric features
overlap.
For the 2025 August and October runs, the typical RMS was $\lesssim$1\%.
In contrast, the 2025 February run was carried out under substantially stronger telluric
absorption, with telluric line depths more than a factor of two larger than in the other runs.
The residuals in the telluric fits for the 2025 February run were therefore higher,
typically around 1.5\,\% and extending to 3--4\,\%, indicating larger post-correction noise.
These larger residuals mainly reflect the difficulty of correcting very strong telluric
absorption rather than a limitation of the PCA framework itself.
All spectra were processed homogeneously for a given instrumental setting, but the
effective noise level after telluric correction still depends on the atmospheric
absorption at the time of observation.

\section{Effective Degrees of Freedom in Residual-Curve Fitting} \label{sec:effective_dof}

In high-resolution spectroscopy, neighboring pixels are not statistically independent.
The instrumental line-spread function, blaze and flat-field corrections, and resampling
onto a common wavelength grid introduce correlations on scales of one to a few pixels,
so the formal number of pixels $N$ overestimates the number of independent data points.
As a result, residual-based fit statistics such as sums of squared residuals cannot be
interpreted using the raw pixel count alone.
This motivates the use of an effective number of independent pixels,
$N_{\mathrm{eff}}$, as discussed in previous studies
\citep[e.g.,][]{Meiksin-2001,Schonebeck-2014}.

Following the standard integrated-autocorrelation approximation, we estimated
$N_{\mathrm{eff}}$ from the autocorrelation of continuum residuals.
For each spectral order, we detrended the continuum with a low-order polynomial,
computed the residual autocorrelations $\rho_j$, and evaluated
\begin{equation}
  N_{\mathrm{eff}}(k) \;=\; \frac{N}{1 + 2 \sum_{j=1}^{k} \rho_j},
\end{equation}
where we adopted $k=10$, which is sufficiently large that $\rho_j$ has decayed to values
consistent with zero within the noise.
Applying this procedure to telluric standard spectra, we found that the terminal values of
$N_{\mathrm{eff}}/N$ cluster around $\sim$0.25 with only modest order-to-order variation.
We therefore adopted $N_{\mathrm{eff}}/N = 0.25$ throughout the abundance analysis.

\section{Behavior of Strong Lines in Solar Analogs} \label{sec:strong_lines}

In the solar-analog sample, we examined how line strength affects
line-by-line abundance residuals prior to applying the per-line calibration.
For each absorption line, we computed $\Delta X=\mathrm{[X/H]}-\mathrm{[X/H]_{lit}}$
for each star as the difference between the uncalibrated line-based abundance estimate
(using only \textit{two-sided} constraints; Section~\ref{subsec:rms_curve})
and the literature abundance \citep{Bedell-2018,Spina-2018}.
For lines with useful $\Delta X$ values in 10 or more stars, we inspected how the
median and RMS scatter of $\Delta X$ vary with line strength.

In contrast to the Cepheid sample studied by \citet{Elgueta-2024}, where strong lines
(depth $>$0.2) showed large systematic offsets ($\gtrsim 0.5$~dex), the depth dependence in the solar analogs
is much weaker. The median $\Delta X$ values vary only mildly across the full range of
line depths, and the RMS scatter remains modest, typically $\sim$0.05--0.15~dex, even for
the strongest lines. This indicates that, within the narrow stellar-parameter range
of solar analogs, abundances inferred from strong lines can remain internally consistent
with those derived from weaker lines.

We therefore retain strong lines in the solar-analog abundance analysis and provide
empirical calibration offsets for them.
Line saturation typically becomes important for equivalent widths around 30--50~m{\AA},
corresponding roughly to {$\sim$}0.1 in depth in normalized solar-analog spectra, but
some lines we use in our abundance measurements are as strong as 200~m{\AA} (depth $\sim$0.4).
In the main analysis, the empirical zero-point correction $\Delta_\mathrm{line}$
(Table~\ref{tab:accepted_lines}) largely absorbs the remaining systematic offsets,
provided that the line strength does not vary strongly across the sample.
Their relatively stable behavior likely reflects the narrow range of atmospheric parameters
in the sample, which limits variations in noncanonical effects such as departures from LTE
or three-dimensional atmospheric structure.
However, strong lines should still be treated with caution in analyses of stars spanning
broader parameter ranges.

\section{Atlas of Combined Reference Spectra and Selected Lines} \label{sec:spectra_with_lines}

Figure~\ref{fig:spectra_with_lines} presents an atlas of the combined, continuum-normalized 
spectrum of the solar-analog calibrators (Section~\ref{subsec:combined_spec}),
with the absorption lines in our final line list (Section~\ref{subsec:line_selection}) overplotted.
The $Y$ and $J$ bands are each divided into 12 panels, covering the full wavelength ranges
used in the abundance analysis (9760--10890\,{\AA}\ and 11820--13190\,{\AA}, respectively).

\begin{figure*}[!tbp]
\centering
\includegraphics[width=0.92\linewidth]{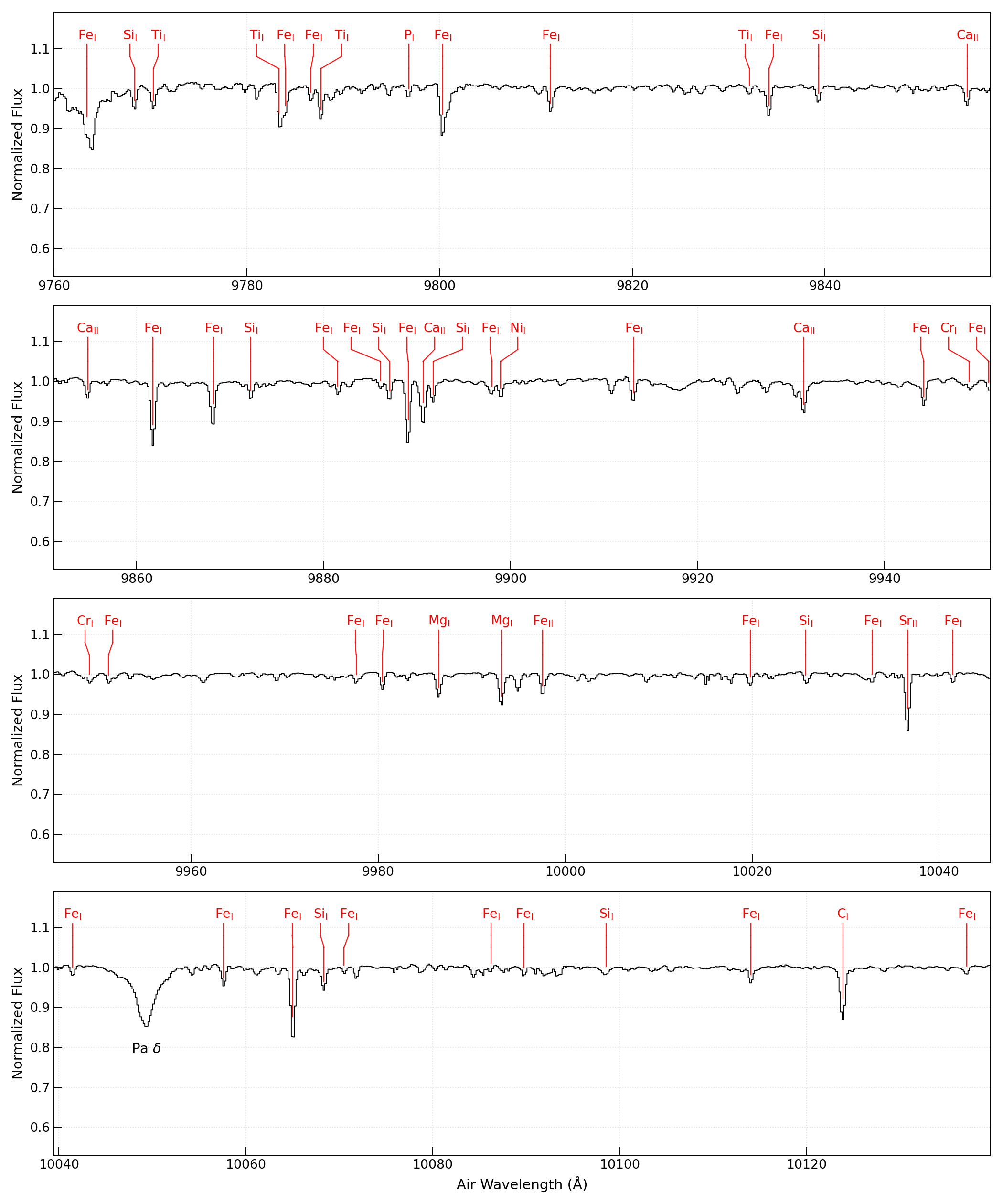}
\caption{Atlas of the combined spectrum of solar-analog calibrators (ages 3--7~Gyr; Section~\ref{subsec:combined_spec}),
shown in air wavelengths and continuum-normalized units.
Red markers indicate accepted lines with species labeled (Table~\ref{tab:accepted_lines}).
The spectrum and the line list are available in the Zenodo repository for this work, \url{https://doi.org/10.5281/zenodo.19230423}. Panels 01--04 in the $Y$ band.}
\label{fig:spectra_with_lines}
\end{figure*}

\addtocounter{figure}{-1}
\begin{figure*}[!tbp]
\centering
\includegraphics[width=0.92\linewidth]{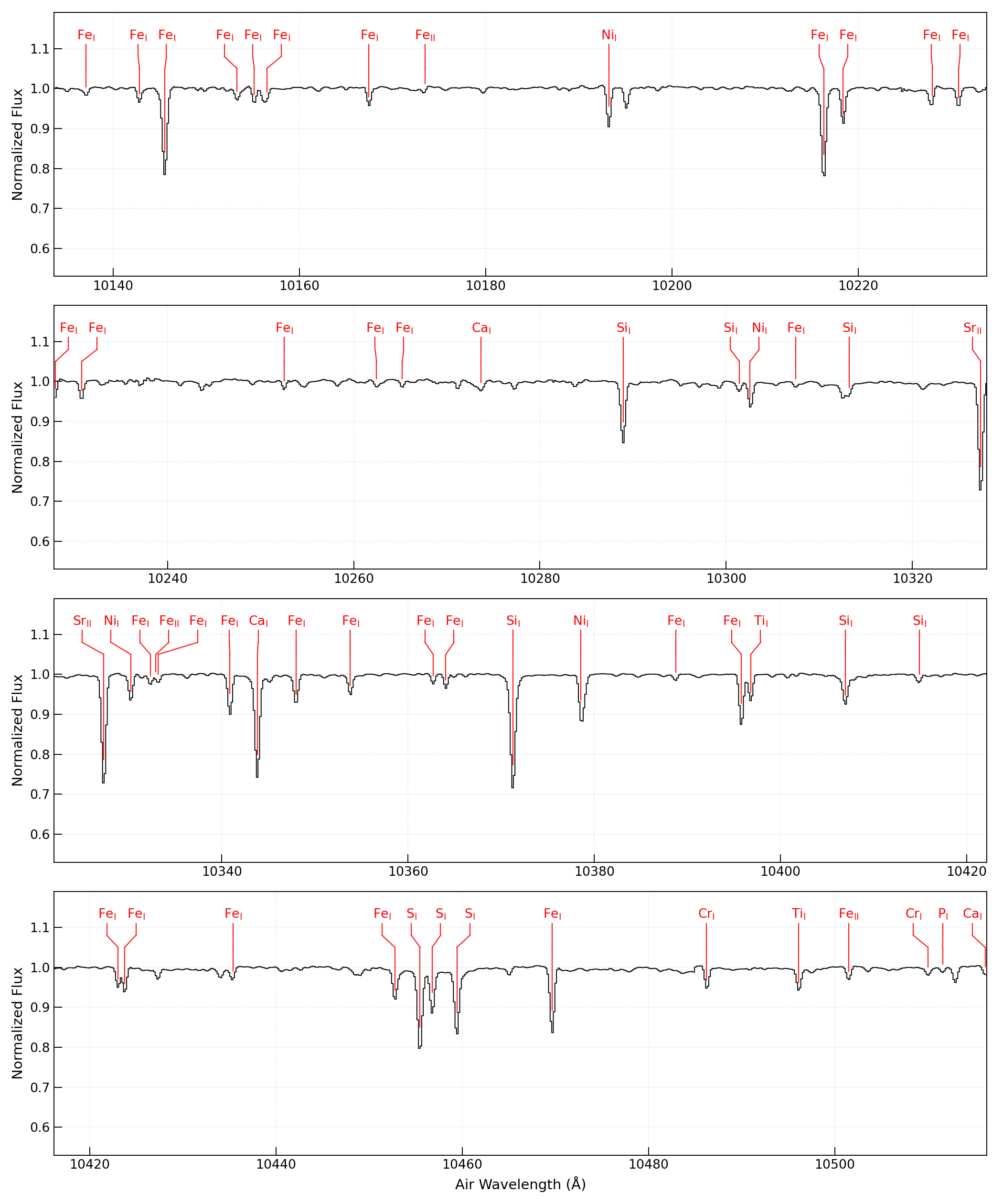}
\caption{---Continued. Panels 05--08 in the $Y$ band.}
\end{figure*}

\addtocounter{figure}{-1}
\begin{figure*}[!tbp]
\centering
\includegraphics[width=0.92\linewidth]{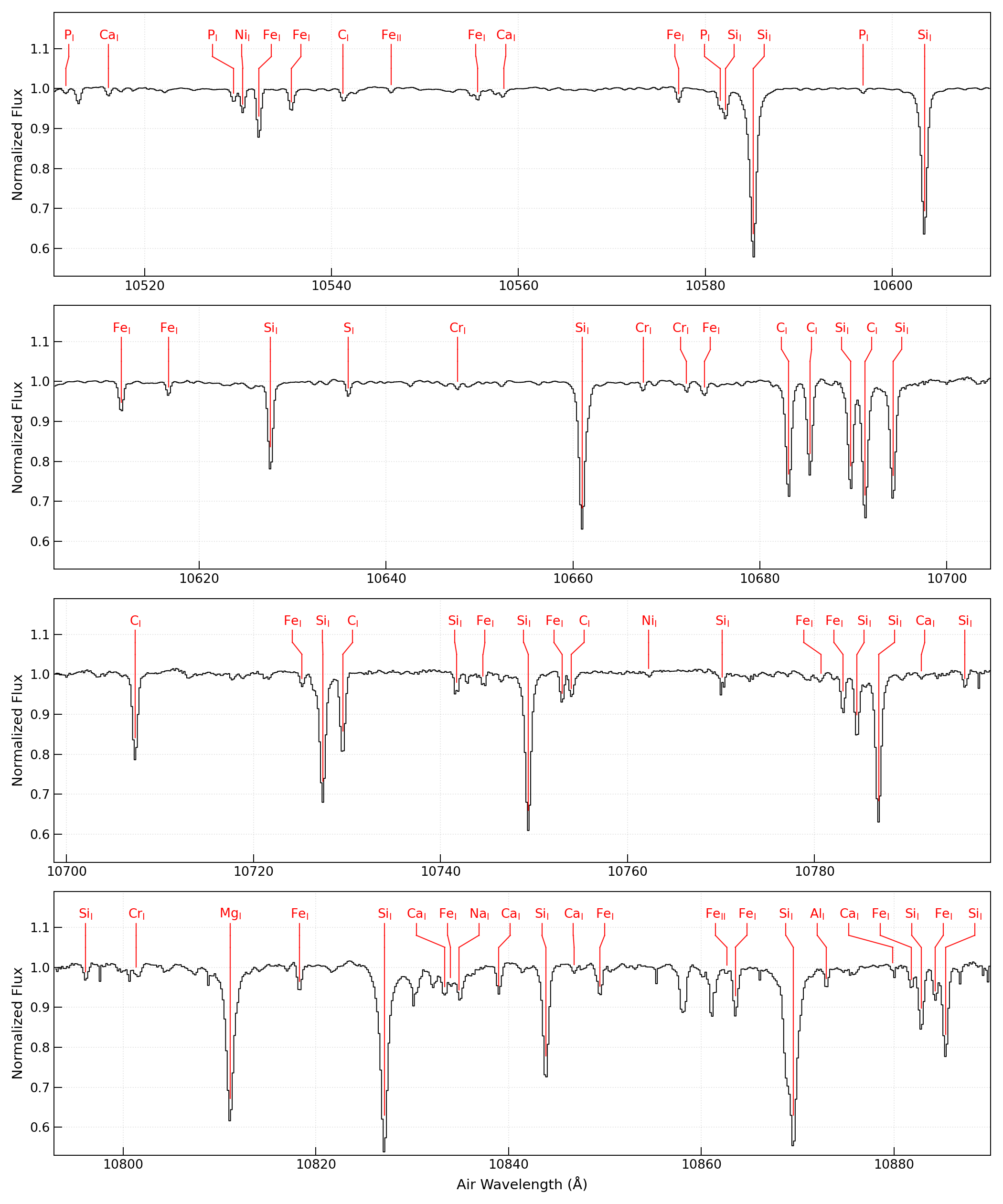}
\caption{---Continued. Panels 09--12 in the $Y$ band.}
\end{figure*}

\addtocounter{figure}{-1}
\begin{figure*}[!tbp]
\centering
\includegraphics[width=0.92\linewidth]{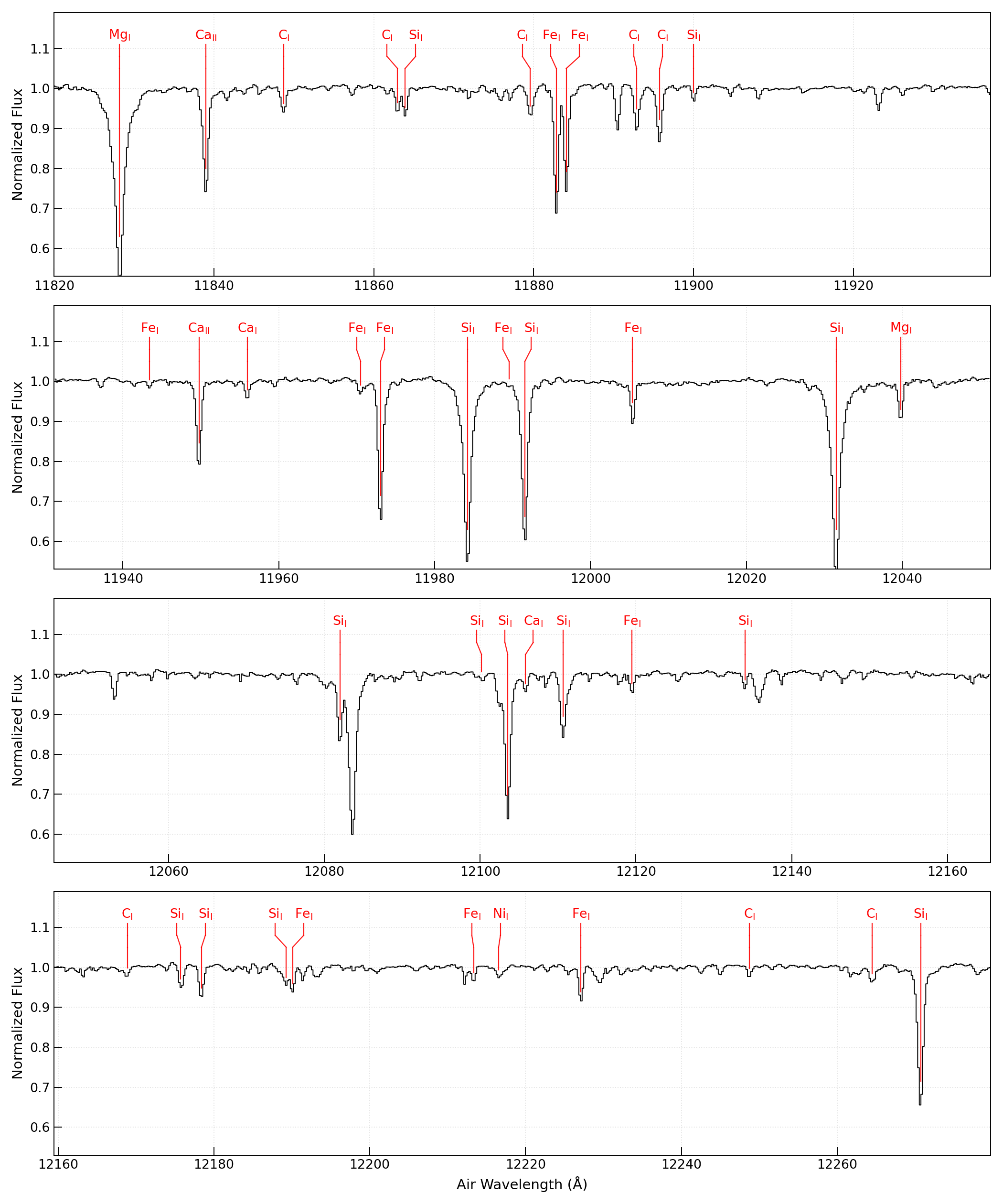}
\caption{---Continued. Panels 01--04 in the $J$ band.}
\end{figure*}

\addtocounter{figure}{-1}
\begin{figure*}[!tbp]
\centering
\includegraphics[width=0.92\linewidth]{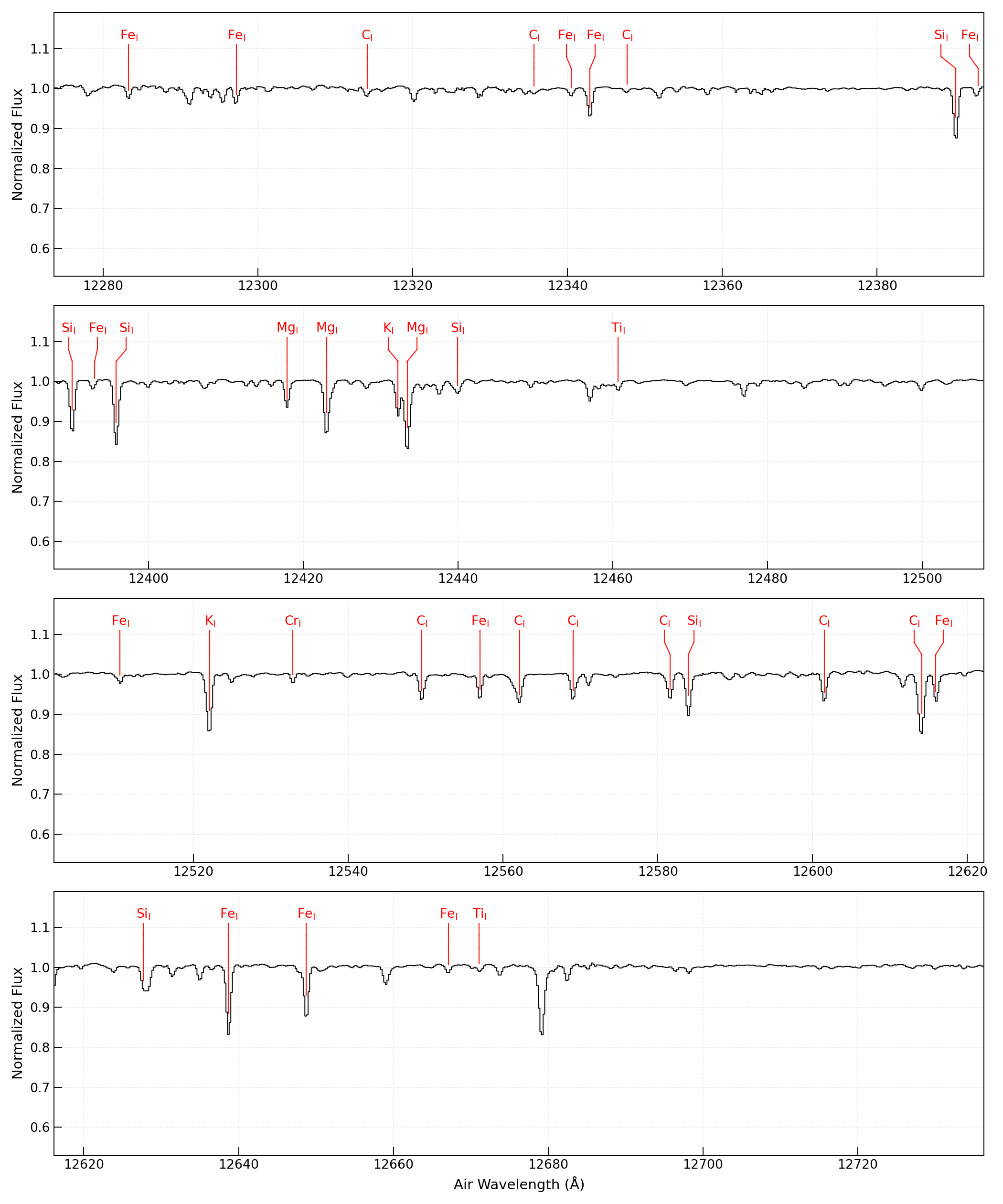}
\caption{---Continued. Panels 05--08 in the $J$ band.}
\end{figure*}

\addtocounter{figure}{-1}
\begin{figure*}[!tbp]
\centering
\includegraphics[width=0.92\linewidth]{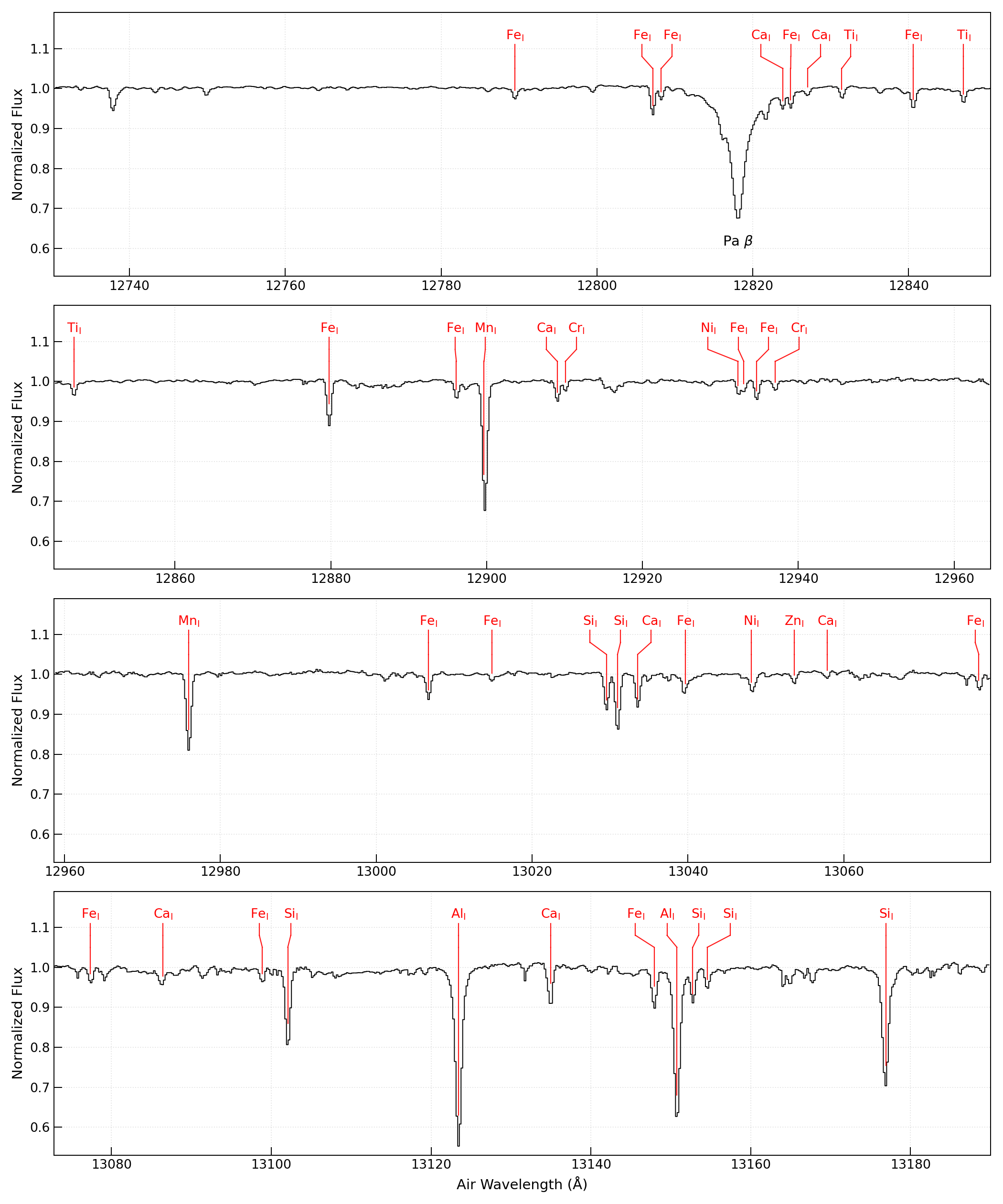}
\caption{---Continued. Panels 09--12 in the $J$ band.}
\end{figure*}

\bibliography{ms}{}

@INPROCEEDINGS{Artigau-2014,
       author = {{Artigau}, {\'E}tienne and {Astudillo-Defru}, Nicola and {Delfosse}, Xavier and {Bouchy}, Fran{\c{c}}ois and {Bonfils}, Xavier and {Lovis}, Christophe and {Pepe}, Francesco and {Moutou}, Claire and {Donati}, Jean-Fran{\c{c}}ois and {Doyon}, Ren{\'e} and {Malo}, Lison},
        title = "{Telluric-line subtraction in high-accuracy velocimetry: a PCA-based approach}",
    booktitle = {Observatory Operations: Strategies, Processes, and Systems V},
         year = 2014,
       editor = {{Peck}, Alison B. and {Benn}, Chris R. and {Seaman}, Robert L.},
       series = {Society of Photo-Optical Instrumentation Engineers (SPIE) Conference Series},
       volume = {9149},
        month = jul,
          eid = {914905},
        pages = {914905},
          doi = {10.1117/12.2056385},
archivePrefix = {arXiv},
       eprint = {1406.6927},
 primaryClass = {astro-ph.IM},
       adsurl = {https://ui.adsabs.harvard.edu/abs/2014SPIE.9149E..05A}
}

@ARTICLE{Aschenbrenner-2025,
       author = {{Aschenbrenner}, P. and {Butler}, K. and {Przybilla}, N.},
        title = "{The present-day cosmic phosphorus abundance}",
      journal = {\aap},
         year = 2025,
        month = jun,
       volume = {698},
          eid = {A164},
        pages = {A164},
          doi = {10.1051/0004-6361/202554356},
archivePrefix = {arXiv},
       eprint = {2505.08428},
 primaryClass = {astro-ph.SR},
       adsurl = {https://ui.adsabs.harvard.edu/abs/2025A&A...698A.164A}
}

@ARTICLE{Asplund-2009,
       author = {{Asplund}, Martin and {Grevesse}, Nicolas and {Sauval}, A. Jacques and {Scott}, Pat},
        title = "{The Chemical Composition of the Sun}",
      journal = {\araa},
         year = 2009,
        month = sep,
       volume = {47},
       number = {1},
        pages = {481-522},
          doi = {10.1146/annurev.astro.46.060407.145222},
archivePrefix = {arXiv},
       eprint = {0909.0948},
 primaryClass = {astro-ph.SR},
       adsurl = {https://ui.adsabs.harvard.edu/abs/2009ARA&A..47..481A}
}

@ARTICLE{astropy-2013,
       author = {{Astropy Collaboration} and {Robitaille}, Thomas P. and {Tollerud}, Erik J. and {Greenfield}, Perry and {Droettboom}, Michael and {Bray}, Erik and {Aldcroft}, Tom and {Davis}, Matt and {Ginsburg}, Adam and {Price-Whelan}, Adrian M. and {Kerzendorf}, Wolfgang E. and {Conley}, Alexander and {Crighton}, Neil and {Barbary}, Kyle and {Muna}, Demitri and {Ferguson}, Henry and {Grollier}, Fr{\'e}d{\'e}ric and {Parikh}, Madhura M. and {Nair}, Prasanth H. and {Unther}, Hans M. and {Deil}, Christoph and {Woillez}, Julien and {Conseil}, Simon and {Kramer}, Roban and {Turner}, James E.~H. and {Singer}, Leo and {Fox}, Ryan and {Weaver}, Benjamin A. and {Zabalza}, Victor and {Edwards}, Zachary I. and {Azalee Bostroem}, K. and {Burke}, D.~J. and {Casey}, Andrew R. and {Crawford}, Steven M. and {Dencheva}, Nadia and {Ely}, Justin and {Jenness}, Tim and {Labrie}, Kathleen and {Lim}, Pey Lian and {Pierfederici}, Francesco and {Pontzen}, Andrew and {Ptak}, Andy and {Refsdal}, Brian and {Servillat}, Mathieu and {Streicher}, Ole},
        title = "{Astropy: A community Python package for astronomy}",
      journal = {\aap},
         year = 2013,
        month = oct,
       volume = {558},
          eid = {A33},
        pages = {A33},
          doi = {10.1051/0004-6361/201322068},
archivePrefix = {arXiv},
       eprint = {1307.6212},
 primaryClass = {astro-ph.IM},
       adsurl = {https://ui.adsabs.harvard.edu/abs/2013A&A...558A..33A}
}

@ARTICLE{astropy-2018,
       author = {{Astropy Collaboration} and {Price-Whelan}, A.~M. and {Sip{\H{o}}cz}, B.~M. and {G{\"u}nther}, H.~M. and {Lim}, P.~L. and {Crawford}, S.~M. and {Conseil}, S. and {Shupe}, D.~L. and {Craig}, M.~W. and {Dencheva}, N. and {Ginsburg}, A. and {VanderPlas}, J.~T. and {Bradley}, L.~D. and {P{\'e}rez-Su{\'a}rez}, D. and {de Val-Borro}, M. and {Aldcroft}, T.~L. and {Cruz}, K.~L. and {Robitaille}, T.~P. and {Tollerud}, E.~J. and {Ardelean}, C. and {Babej}, T. and {Bach}, Y.~P. and {Bachetti}, M. and {Bakanov}, A.~V. and {Bamford}, S.~P. and {Barentsen}, G. and {Barmby}, P. and {Baumbach}, A. and {Berry}, K.~L. and {Biscani}, F. and {Boquien}, M. and {Bostroem}, K.~A. and {Bouma}, L.~G. and {Brammer}, G.~B. and {Bray}, E.~M. and {Breytenbach}, H. and {Buddelmeijer}, H. and {Burke}, D.~J. and {Calderone}, G. and {Cano Rodr{\'\i}guez}, J.~L. and {Cara}, M. and {Cardoso}, J.~V.~M. and {Cheedella}, S. and {Copin}, Y. and {Corrales}, L. and {Crichton}, D. and {D'Avella}, D. and {Deil}, C. and {Depagne}, {\'E}. and {Dietrich}, J.~P. and {Donath}, A. and {Droettboom}, M. and {Earl}, N. and {Erben}, T. and {Fabbro}, S. and {Ferreira}, L.~A. and {Finethy}, T. and {Fox}, R.~T. and {Garrison}, L.~H. and {Gibbons}, S.~L.~J. and {Goldstein}, D.~A. and {Gommers}, R. and {Greco}, J.~P. and {Greenfield}, P. and {Groener}, A.~M. and {Grollier}, F. and {Hagen}, A. and {Hirst}, P. and {Homeier}, D. and {Horton}, A.~J. and {Hosseinzadeh}, G. and {Hu}, L. and {Hunkeler}, J.~S. and {Ivezi{\'c}}, {\v{Z}}. and {Jain}, A. and {Jenness}, T. and {Kanarek}, G. and {Kendrew}, S. and {Kern}, N.~S. and {Kerzendorf}, W.~E. and {Khvalko}, A. and {King}, J. and {Kirkby}, D. and {Kulkarni}, A.~M. and {Kumar}, A. and {Lee}, A. and {Lenz}, D. and {Littlefair}, S.~P. and {Ma}, Z. and {Macleod}, D.~M. and {Mastropietro}, M. and {McCully}, C. and {Montagnac}, S. and {Morris}, B.~M. and {Mueller}, M. and {Mumford}, S.~J. and {Muna}, D. and {Murphy}, N.~A. and {Nelson}, S. and {Nguyen}, G.~H. and {Ninan}, J.~P. and {N{\"o}the}, M. and {Ogaz}, S. and {Oh}, S. and {Parejko}, J.~K. and {Parley}, N. and {Pascual}, S. and {Patil}, R. and {Patil}, A.~A. and {Plunkett}, A.~L. and {Prochaska}, J.~X. and {Rastogi}, T. and {Reddy Janga}, V. and {Sabater}, J. and {Sakurikar}, P. and {Seifert}, M. and {Sherbert}, L.~E. and {Sherwood-Taylor}, H. and {Shih}, A.~Y. and {Sick}, J. and {Silbiger}, M.~T. and {Singanamalla}, S. and {Singer}, L.~P. and {Sladen}, P.~H. and {Sooley}, K.~A. and {Sornarajah}, S. and {Streicher}, O. and {Teuben}, P. and {Thomas}, S.~W. and {Tremblay}, G.~R. and {Turner}, J.~E.~H. and {Terr{\'o}n}, V. and {van Kerkwijk}, M.~H. and {de la Vega}, A. and {Watkins}, L.~L. and {Weaver}, B.~A. and {Whitmore}, J.~B. and {Woillez}, J. and {Zabalza}, V. and {Astropy Contributors}},
        title = "{The Astropy Project: Building an Open-science Project and Status of the v2.0 Core Package}",
      journal = {\aj},
         year = 2018,
        month = sep,
       volume = {156},
       number = {3},
          eid = {123},
        pages = {123},
          doi = {10.3847/1538-3881/aabc4f},
archivePrefix = {arXiv},
       eprint = {1801.02634},
 primaryClass = {astro-ph.IM},
       adsurl = {https://ui.adsabs.harvard.edu/abs/2018AJ....156..123A}
}

@ARTICLE{astropy-2022,
       author = {{Astropy Collaboration} and {Price-Whelan}, Adrian M. and {Lim}, Pey Lian and {Earl}, Nicholas and {Starkman}, Nathaniel and {Bradley}, Larry and {Shupe}, David L. and {Patil}, Aarya A. and {Corrales}, Lia and {Brasseur}, C.~E. and {N{\"o}the}, Maximilian and {Donath}, Axel and {Tollerud}, Erik and {Morris}, Brett M. and {Ginsburg}, Adam and {Vaher}, Eero and {Weaver}, Benjamin A. and {Tocknell}, James and {Jamieson}, William and {van Kerkwijk}, Marten H. and {Robitaille}, Thomas P. and {Merry}, Bruce and {Bachetti}, Matteo and {G{\"u}nther}, H. Moritz and {Aldcroft}, Thomas L. and {Alvarado-Montes}, Jaime A. and {Archibald}, Anne M. and {B{\'o}di}, Attila and {Bapat}, Shreyas and {Barentsen}, Geert and {Baz{\'a}n}, Juanjo and {Biswas}, Manish and {Boquien}, M{\'e}d{\'e}ric and {Burke}, D.~J. and {Cara}, Daria and {Cara}, Mihai and {Conroy}, Kyle E. and {Conseil}, Simon and {Craig}, Matthew W. and {Cross}, Robert M. and {Cruz}, Kelle L. and {D'Eugenio}, Francesco and {Dencheva}, Nadia and {Devillepoix}, Hadrien A.~R. and {Dietrich}, J{\"o}rg P. and {Eigenbrot}, Arthur Davis and {Erben}, Thomas and {Ferreira}, Leonardo and {Foreman-Mackey}, Daniel and {Fox}, Ryan and {Freij}, Nabil and {Garg}, Suyog and {Geda}, Robel and {Glattly}, Lauren and {Gondhalekar}, Yash and {Gordon}, Karl D. and {Grant}, David and {Greenfield}, Perry and {Groener}, Austen M. and {Guest}, Steve and {Gurovich}, Sebastian and {Handberg}, Rasmus and {Hart}, Akeem and {Hatfield-Dodds}, Zac and {Homeier}, Derek and {Hosseinzadeh}, Griffin and {Jenness}, Tim and {Jones}, Craig K. and {Joseph}, Prajwel and {Kalmbach}, J. Bryce and {Karamehmetoglu}, Emir and {Ka{\l}uszy{\'n}ski}, Miko{\l}aj and {Kelley}, Michael S.~P. and {Kern}, Nicholas and {Kerzendorf}, Wolfgang E. and {Koch}, Eric W. and {Kulumani}, Shankar and {Lee}, Antony and {Ly}, Chun and {Ma}, Zhiyuan and {MacBride}, Conor and {Maljaars}, Jakob M. and {Muna}, Demitri and {Murphy}, N.~A. and {Norman}, Henrik and {O'Steen}, Richard and {Oman}, Kyle A. and {Pacifici}, Camilla and {Pascual}, Sergio and {Pascual-Granado}, J. and {Patil}, Rohit R. and {Perren}, Gabriel I. and {Pickering}, Timothy E. and {Rastogi}, Tanuj and {Roulston}, Benjamin R. and {Ryan}, Daniel F. and {Rykoff}, Eli S. and {Sabater}, Jose and {Sakurikar}, Parikshit and {Salgado}, Jes{\'u}s and {Sanghi}, Aniket and {Saunders}, Nicholas and {Savchenko}, Volodymyr and {Schwardt}, Ludwig and {Seifert-Eckert}, Michael and {Shih}, Albert Y. and {Jain}, Anany Shrey and {Shukla}, Gyanendra and {Sick}, Jonathan and {Simpson}, Chris and {Singanamalla}, Sudheesh and {Singer}, Leo P. and {Singhal}, Jaladh and {Sinha}, Manodeep and {Sip{\H{o}}cz}, Brigitta M. and {Spitler}, Lee R. and {Stansby}, David and {Streicher}, Ole and {{\v{S}}umak}, Jani and {Swinbank}, John D. and {Taranu}, Dan S. and {Tewary}, Nikita and {Tremblay}, Grant R. and {de Val-Borro}, Miguel and {Van Kooten}, Samuel J. and {Vasovi{\'c}}, Zlatan and {Verma}, Shresth and {de Miranda Cardoso}, Jos{\'e} Vin{\'\i}cius and {Williams}, Peter K.~G. and {Wilson}, Tom J. and {Winkel}, Benjamin and {Wood-Vasey}, W.~M. and {Xue}, Rui and {Yoachim}, Peter and {Zhang}, Chen and {Zonca}, Andrea and {Astropy Project Contributors}},
        title = "{The Astropy Project: Sustaining and Growing a Community-oriented Open-source Project and the Latest Major Release (v5.0) of the Core Package}",
      journal = {\apj},
         year = 2022,
        month = aug,
       volume = {935},
       number = {2},
          eid = {167},
        pages = {167},
          doi = {10.3847/1538-4357/ac7c74},
archivePrefix = {arXiv},
       eprint = {2206.14220},
 primaryClass = {astro-ph.IM},
       adsurl = {https://ui.adsabs.harvard.edu/abs/2022ApJ...935..167A}
}

@ARTICLE{Bedell-2018,
       author = {{Bedell}, Megan and {Bean}, Jacob L. and {Mel{\'e}ndez}, Jorge and {Spina}, Lorenzo and {Ram{\'\i}rez}, Ivan and {Asplund}, Martin and {Alves-Brito}, Alan and {dos Santos}, Leonardo and {Dreizler}, Stefan and {Yong}, David and {Monroe}, TalaWanda and {Casagrande}, Luca},
        title = "{The Chemical Homogeneity of Sun-like Stars in the Solar Neighborhood}",
      journal = {\apj},
         year = 2018,
        month = sep,
       volume = {865},
       number = {1},
          eid = {68},
        pages = {68},
          doi = {10.3847/1538-4357/aad908},
archivePrefix = {arXiv},
       eprint = {1802.02576},
 primaryClass = {astro-ph.SR},
       adsurl = {https://ui.adsabs.harvard.edu/abs/2018ApJ...865...68B}
}

@ARTICLE{Bekki-2024,
       author = {{Bekki}, Kenji and {Tsujimoto}, Takuji},
        title = "{Phosphorus Enrichment by ONe Novae in the Galaxy}",
      journal = {\apjl},
         year = 2024,
        month = may,
       volume = {967},
       number = {1},
          eid = {L1},
        pages = {L1},
          doi = {10.3847/2041-8213/ad3fb6},
archivePrefix = {arXiv},
       eprint = {2405.06325},
 primaryClass = {astro-ph.GA},
       adsurl = {https://ui.adsabs.harvard.edu/abs/2024ApJ...967L...1B}
}

@ARTICLE{Blackwell-Whitehead-2011,
       author = {{Blackwell-Whitehead}, R. and {Pavlenko}, Y.~V. and {Nave}, G. and {Pickering}, J.~C. and {Jones}, H.~R.~A. and {Lyubchik}, Y. and {Nilsson}, H.},
        title = "{Infrared Mn I laboratory oscillator strengths for the study of late type stars and ultracool dwarfs}",
      journal = {\aap},
         year = 2011,
        month = jan,
       volume = {525},
          eid = {A44},
        pages = {A44},
          doi = {10.1051/0004-6361/201015006},
       adsurl = {https://ui.adsabs.harvard.edu/abs/2011A&A...525A..44B}
}

@ARTICLE{Caffau-2011,
       author = {{Caffau}, E. and {Bonifacio}, P. and {Faraggiana}, R. and {Steffen}, M.},
        title = "{The Galactic evolution of phosphorus}",
      journal = {\aap},
         year = 2011,
        month = aug,
       volume = {532},
          eid = {A98},
        pages = {A98},
          doi = {10.1051/0004-6361/201117313},
archivePrefix = {arXiv},
       eprint = {1107.2657},
 primaryClass = {astro-ph.GA},
       adsurl = {https://ui.adsabs.harvard.edu/abs/2011A&A...532A..98C}
}

@ARTICLE{Caffau-2016,
       author = {{Caffau}, E. and {Andrievsky}, S. and {Korotin}, S. and {Origlia}, L. and {Oliva}, E. and {Sanna}, N. and {Ludwig}, H. -G. and {Bonifacio}, P.},
        title = "{GIANO Y-band spectroscopy of dwarf stars: Phosphorus, sulphur, and strontium abundances}",
      journal = {\aap},
         year = 2016,
        month = jan,
       volume = {585},
          eid = {A16},
        pages = {A16},
          doi = {10.1051/0004-6361/201527272},
archivePrefix = {arXiv},
       eprint = {1510.06396},
 primaryClass = {astro-ph.SR},
       adsurl = {https://ui.adsabs.harvard.edu/abs/2016A&A...585A..16C}
}

@ARTICLE{Caffau-2019,
       author = {{Caffau}, E. and {Bonifacio}, P. and {Oliva}, E. and {Korotin}, S. and {Capitanio}, L. and {Andrievsky}, S. and {Collet}, R. and {Sbordone}, L. and {Duffau}, S. and {Sanna}, N. and {Tozzi}, A. and {Origlia}, L. and {Ryde}, N. and {Ludwig}, H. -G.},
        title = "{Systematic investigation of chemical abundances derived using IR spectra obtained with GIANO}",
      journal = {\aap},
         year = 2019,
        month = feb,
       volume = {622},
          eid = {A68},
        pages = {A68},
          doi = {10.1051/0004-6361/201834318},
archivePrefix = {arXiv},
       eprint = {1812.05100},
 primaryClass = {astro-ph.SR},
       adsurl = {https://ui.adsabs.harvard.edu/abs/2019A&A...622A..68C}
}

@ARTICLE{Cescutti-2012,
       author = {{Cescutti}, G. and {Matteucci}, F. and {Caffau}, E. and {Fran{\c{c}}ois}, P.},
        title = "{Chemical evolution of the Milky Way: the origin of phosphorus}",
      journal = {\aap},
         year = 2012,
        month = apr,
       volume = {540},
          eid = {A33},
        pages = {A33},
          doi = {10.1051/0004-6361/201118188},
archivePrefix = {arXiv},
       eprint = {1112.3824},
 primaryClass = {astro-ph.GA},
       adsurl = {https://ui.adsabs.harvard.edu/abs/2012A&A...540A..33C}
}

@ARTICLE{DelgadoMena-2019,
       author = {{Delgado Mena}, E. and {Moya}, A. and {Adibekyan}, V. and {Tsantaki}, M. and {Gonz{\'a}lez Hern{\'a}ndez}, J.~I. and {Israelian}, G. and {Davies}, G.~R. and {Chaplin}, W.~J. and {Sousa}, S.~G. and {Ferreira}, A.~C.~S. and {Santos}, N.~C.},
        title = "{Abundance to age ratios in the HARPS-GTO sample with Gaia DR2. Chemical clocks for a range of [Fe/H]}",
      journal = {\aap},
         year = 2019,
        month = apr,
       volume = {624},
          eid = {A78},
        pages = {A78},
          doi = {10.1051/0004-6361/201834783},
archivePrefix = {arXiv},
       eprint = {1902.02127},
 primaryClass = {astro-ph.SR},
       adsurl = {https://ui.adsabs.harvard.edu/abs/2019A&A...624A..78D}
}

@ARTICLE{Elgueta-2024,
       author = {{Elgueta}, S.~S. and {Matsunaga}, N. and {Jian}, M. and {Taniguchi}, D. and {Kobayashi}, N. and {Fukue}, K. and {Hamano}, S. and {Sameshima}, H. and {Kondo}, S. and {Arai}, A. and {Ikeda}, Y. and {Kawakita}, H. and {Otsubo}, S. and {Yasui}, C. and {Tsujimoto}, T.},
        title = "{Astrophysical calibration of the oscillator strengths of YJ-band absorption lines in classical Cepheids}",
      journal = {\mnras},
         year = 2024,
        month = aug,
       volume = {532},
       number = {4},
        pages = {3694-3712},
          doi = {10.1093/mnras/stae1674},
archivePrefix = {arXiv},
       eprint = {2307.00158},
 primaryClass = {astro-ph.SR},
       adsurl = {https://ui.adsabs.harvard.edu/abs/2024MNRAS.532.3694E}
}

@ARTICLE{Fukue-2021,
       author = {{Fukue}, Kei and {Matsunaga}, Noriyuki and {Kondo}, Sohei and {Taniguchi}, Daisuke and {Ikeda}, Yuji and {Kobayashi}, Naoto and {Sameshima}, Hiroaki and {Hamano}, Satoshi and {Arai}, Akira and {Kawakita}, Hideyo and {Yasui}, Chikako and {Mizumoto}, Misaki and {Otsubo}, Shogo and {Takenaka}, Keiichi and {Yoshikawa}, Tomohiro and {Tsujimoto}, Takuji},
        title = "{Absorption Lines in the 0.91-1.33 {\ensuremath{\mu}}m Spectra of Red Giants for Measuring Abundances of Mg, Si, Ca, Ti, Cr, and Ni}",
      journal = {\apj},
         year = 2021,
        month = may,
       volume = {913},
       number = {1},
          eid = {62},
        pages = {62},
          doi = {10.3847/1538-4357/abf0b1},
archivePrefix = {arXiv},
       eprint = {2103.12478},
 primaryClass = {astro-ph.SR},
       adsurl = {https://ui.adsabs.harvard.edu/abs/2021ApJ...913...62F}
}

@ARTICLE{Gratton-2006,
       author = {{Gratton}, Raffaele and {Bragaglia}, Angela and {Carretta}, Eugenio and {Tosi}, Monica},
        title = "{The Metallicity of the Old Open Cluster NGC 6791}",
      journal = {\apj},
         year = 2006,
        month = may,
       volume = {642},
       number = {1},
        pages = {462-469},
          doi = {10.1086/500729},
archivePrefix = {arXiv},
       eprint = {astro-ph/0601027},
 primaryClass = {astro-ph},
       adsurl = {https://ui.adsabs.harvard.edu/abs/2006ApJ...642..462G}
}

@ARTICLE{Hamano-2024,
       author = {{Hamano}, Satoshi and {Ikeda}, Yuji and {Otsubo}, Shogo and {Katoh}, Haruki and {Fukue}, Kei and {Matsunaga}, Noriyuki and {Taniguchi}, Daisuke and {Kawakita}, Hideyo and {Takenaka}, Keiichi and {Kondo}, Sohei and {Sameshima}, Hiroaki},
        title = "{WARP: The Data Reduction Pipeline for the WINERED Spectrograph}",
      journal = {\pasp},
         year = 2024,
        month = jan,
       volume = {136},
       number = {1},
          eid = {014504},
        pages = {014504},
          doi = {10.1088/1538-3873/ad1b38},
archivePrefix = {arXiv},
       eprint = {2401.04876},
 primaryClass = {astro-ph.IM},
       adsurl = {https://ui.adsabs.harvard.edu/abs/2024PASP..136a4504H}
}

@INPROCEEDINGS{Ikeda-2018,
       author = {{Ikeda}, Yuji and {Kobayashi}, Naoto and {Kondo}, Sohei and {Otsubo}, Shogo and {Watase}, Ayaka and {Murai}, Taichi and {Sakamoto}, Kyoko and {Hamano}, Satoshi and {Sameshima}, Hiroaki and {Fukue}, Kei and {Arai}, Akira and {Yoshikawa}, Tomohiro and {Takenaka}, Kei-ichi and {Manome}, Takeo and {Mukai}, Shinji and {Iida}, Naoto and {Yanagibashi}, Kentaro and {Yasui}, Chikako and {Mizumoto}, Misaki and {Matsunaga}, Noriyuki and {Bono}, Giuseppe and {Kawakita}, Hideyo},
        title = "{Very high-sensitive NIR high-resolution spectrograph WINERED: on-going observations at NTT}",
    booktitle = {Ground-based and Airborne Instrumentation for Astronomy VII},
         year = 2018,
       editor = {{Evans}, Christopher J. and {Simard}, Luc and {Takami}, Hideki},
       series = {Society of Photo-Optical Instrumentation Engineers (SPIE) Conference Series},
       volume = {10702},
        month = jul,
          eid = {107025U},
        pages = {107025U},
          doi = {10.1117/12.2309605},
       adsurl = {https://ui.adsabs.harvard.edu/abs/2018SPIE10702E..5UI}
}

@ARTICLE{Ikeda-2022,
       author = {{Ikeda}, Yuji and {Kondo}, Sohei and {Otsubo}, Shogo and {Hamano}, Satoshi and {Yasui}, Chikako and {Matsunaga}, Noriyuki and {Sameshima}, Hiroaki and {Yoshikawa}, Tomohiro and {Fukue}, Kei and {Nakanishi}, Kenshi and {Kawanishi}, Takafumi and {Watase}, Ayaka and {Nakaoka}, Tetsuya and {Arai}, Akira and {Kinoshita}, Masaomi and {Kitano}, Ayaka and {Nakamura}, Kazuki and {Asano}, Akira and {Takenaka}, Keiichi and {Murai}, Taichi and {Kawakita}, Hideyo and {Minami}, Atsushi and {Izumi}, Natsuko and {Yamamoto}, Ryo and {Mizumoto}, Misaki and {Taniguchi}, Daisuke and {Tsujimoto}, Takuji},
        title = "{Highly Sensitive, Non-cryogenic NIR High-resolution Spectrograph, WINERED}",
      journal = {\pasp},
         year = 2022,
        month = jan,
       volume = {134},
       number = {1031},
          eid = {015004},
        pages = {015004},
          doi = {10.1088/1538-3873/ac1c5f},
       adsurl = {https://ui.adsabs.harvard.edu/abs/2022PASP..134a5004I}
}

@ARTICLE{Iwamoto-1999,
       author = {{Iwamoto}, Koichi and {Brachwitz}, Franziska and {Nomoto}, Ken'ICHI and {Kishimoto}, Nobuhiro and {Umeda}, Hideyuki and {Hix}, W. Raphael and {Thielemann}, Friedrich-Karl},
        title = "{Nucleosynthesis in Chandrasekhar Mass Models for Type IA Supernovae and Constraints on Progenitor Systems and Burning-Front Propagation}",
      journal = {\apjs},
         year = 1999,
        month = dec,
       volume = {125},
       number = {2},
        pages = {439-462},
          doi = {10.1086/313278},
archivePrefix = {arXiv},
       eprint = {astro-ph/0002337},
 primaryClass = {astro-ph},
       adsurl = {https://ui.adsabs.harvard.edu/abs/1999ApJS..125..439I}
}

@ARTICLE{Jacobson-2014,
       author = {{Jacobson}, Heather R. and {Thanathibodee}, Thanawuth and {Frebel}, Anna and {Roederer}, Ian U. and {Cescutti}, Gabriele and {Matteucci}, Francesca},
        title = "{The Chemical Evolution of Phosphorus}",
      journal = {\apjl},
         year = 2014,
        month = dec,
       volume = {796},
       number = {2},
          eid = {L24},
        pages = {L24},
          doi = {10.1088/2041-8205/796/2/L24},
archivePrefix = {arXiv},
       eprint = {1410.8538},
 primaryClass = {astro-ph.SR},
       adsurl = {https://ui.adsabs.harvard.edu/abs/2014ApJ...796L..24J}
}

@ARTICLE{Jian-2026,
       author = {{Jian}, Mingjie and {Fu}, Xiaoting and {D'Orazi}, Valentina and {Bragaglia}, Angela and {Bijavara Seshashayana}, S. and {Zhao}, He and {Guo}, Ziyi and {Lind}, Karin and {Matsunaga}, Noriyuki and {Nunnari}, Antonino and {Bono}, Giuseppe and {Sanna}, Nicoletta and {Romano}, Donatella and {Dal Ponte}, Marina},
        title = "{Stellar population astrophysics (SPA) with the TNG. The phosphorus abundance on the young side of Milky Way}",
      journal = {\mnras},
         year = 2026,
        month = jan,
       volume = {545},
       number = {2},
          eid = {staf1797},
        pages = {staf1797},
          doi = {10.1093/mnras/staf1797},
archivePrefix = {arXiv},
       eprint = {2510.14791},
 primaryClass = {astro-ph.SR},
       adsurl = {https://ui.adsabs.harvard.edu/abs/2026MNRAS.545f1797J}
}

@ARTICLE{Jose-1998,
       author = {{Jos{\'e}}, Jordi and {Hernanz}, Margarita},
        title = "{Nucleosynthesis in Classical Novae: CO versus ONe White Dwarfs}",
      journal = {\apj},
         year = 1998,
        month = feb,
       volume = {494},
       number = {2},
        pages = {680-690},
          doi = {10.1086/305244},
archivePrefix = {arXiv},
       eprint = {astro-ph/9709153},
 primaryClass = {astro-ph},
       adsurl = {https://ui.adsabs.harvard.edu/abs/1998ApJ...494..680J}
}

@ARTICLE{Jose-2001,
       author = {{Jos{\'e}}, Jordi and {Coc}, Alain and {Hernanz}, Margarita},
        title = "{Synthesis of Intermediate-Mass Elements in Classical Novae: From Si to Ca}",
      journal = {\apj},
         year = 2001,
        month = oct,
       volume = {560},
       number = {2},
        pages = {897-906},
          doi = {10.1086/322979},
archivePrefix = {arXiv},
       eprint = {astro-ph/0106418},
 primaryClass = {astro-ph},
       adsurl = {https://ui.adsabs.harvard.edu/abs/2001ApJ...560..897J}
}

@ARTICLE{Karakas-2014,
       author = {{Karakas}, Amanda I. and {Lattanzio}, John C.},
        title = "{The Dawes Review 2: Nucleosynthesis and Stellar Yields of Low- and Intermediate-Mass Single Stars}",
      journal = {\pasa},
         year = 2014,
        month = jul,
       volume = {31},
          eid = {e030},
        pages = {e030},
          doi = {10.1017/pasa.2014.21},
archivePrefix = {arXiv},
       eprint = {1405.0062},
 primaryClass = {astro-ph.SR},
       adsurl = {https://ui.adsabs.harvard.edu/abs/2014PASA...31...30K}
}

@ARTICLE{Karakas-2016,
       author = {{Karakas}, Amanda I. and {Lugaro}, Maria},
        title = "{Stellar Yields from Metal-rich Asymptotic Giant Branch Models}",
      journal = {\apj},
         year = 2016,
        month = jul,
       volume = {825},
       number = {1},
          eid = {26},
        pages = {26},
          doi = {10.3847/0004-637X/825/1/26},
archivePrefix = {arXiv},
       eprint = {1604.02178},
 primaryClass = {astro-ph.SR},
       adsurl = {https://ui.adsabs.harvard.edu/abs/2016ApJ...825...26K}
}

@ARTICLE{Kato-1996,
       author = {{Kato}, Ken-Ichi and {Watanabe}, Yoshiya and {Sadakane}, Kozo},
        title = "{Atmospheric Abundances of Light Elements in the F-Type Star Procyon}",
      journal = {\pasj},
         year = 1996,
        month = aug,
       volume = {48},
        pages = {601-606},
          doi = {10.1093/pasj/48.4.601},
       adsurl = {https://ui.adsabs.harvard.edu/abs/1996PASJ...48..601K}
}

@ARTICLE{Kemp-2024,
       author = {{Kemp}, Alex J. and {Karakas}, Amanda I. and {Casey}, Andrew R. and {C{\^o}t{\'e}}, Benoit and {Izzard}, Robert G. and {Osborn}, Zara},
        title = "{Nova contributions to the chemical evolution of the Milky Way}",
      journal = {\aap},
         year = 2024,
        month = sep,
       volume = {689},
          eid = {A222},
        pages = {A222},
          doi = {10.1051/0004-6361/202450800},
archivePrefix = {arXiv},
       eprint = {2407.18718},
 primaryClass = {astro-ph.GA},
       adsurl = {https://ui.adsabs.harvard.edu/abs/2024A&A...689A.222K}
}

@ARTICLE{Kobayashi-2020,
       author = {{Kobayashi}, Chiaki and {Karakas}, Amanda I. and {Lugaro}, Maria},
        title = "{The Origin of Elements from Carbon to Uranium}",
      journal = {\apj},
         year = 2020,
        month = sep,
       volume = {900},
       number = {2},
          eid = {179},
        pages = {179},
          doi = {10.3847/1538-4357/abae65},
archivePrefix = {arXiv},
       eprint = {2008.04660},
 primaryClass = {astro-ph.GA},
       adsurl = {https://ui.adsabs.harvard.edu/abs/2020ApJ...900..179K}
}

@ARTICLE{Kondo-2019,
       author = {{Kondo}, Sohei and {Fukue}, Kei and {Matsunaga}, Noriyuki and {Ikeda}, Yuji and {Taniguchi}, Daisuke and {Kobayashi}, Naoto and {Sameshima}, Hiroaki and {Hamano}, Satoshi and {Arai}, Akira and {Kawakita}, Hideyo and {Yasui}, Chikako and {Izumi}, Natsuko and {Mizumoto}, Misaki and {Otsubo}, Shogo and {Takenaka}, Keiichi and {Watase}, Ayaka and {Asano}, Akira and {Yoshikawa}, Tomohiro and {Tsujimoto}, Takuji},
        title = "{Fe I Lines in 0.91-1.33 {\ensuremath{\mu}}m Spectra of Red Giants for Measuring the Microturbulence and Metallicities}",
      journal = {\apj},
         year = 2019,
        month = apr,
       volume = {875},
       number = {2},
          eid = {129},
        pages = {129},
          doi = {10.3847/1538-4357/ab0ec4},
archivePrefix = {arXiv},
       eprint = {1903.02241},
 primaryClass = {astro-ph.SR},
       adsurl = {https://ui.adsabs.harvard.edu/abs/2019ApJ...875..129K}
}

@ARTICLE{Maas-2022,
       author = {{Maas}, Zachary G. and {Hawkins}, Keith and {Hinkel}, Natalie R. and {Cargile}, Phillip and {Janowiecki}, Steven and {Nelson}, Tyler},
        title = "{The Galactic Distribution of Phosphorus: A Survey of 163 Disk and Halo Stars}",
      journal = {\aj},
         year = 2022,
        month = aug,
       volume = {164},
       number = {2},
          eid = {61},
        pages = {61},
          doi = {10.3847/1538-3881/ac77f8},
archivePrefix = {arXiv},
       eprint = {2206.03528},
 primaryClass = {astro-ph.SR},
       adsurl = {https://ui.adsabs.harvard.edu/abs/2022AJ....164...61M}
}

@ARTICLE{Magain-1984,
       author = {{Magain}, P.},
        title = "{A comment on systematic errors in determinations of microturbulent velocities}",
      journal = {\aap},
         year = 1984,
        month = may,
       volume = {134},
       number = {1},
        pages = {189-192},
       adsurl = {https://ui.adsabs.harvard.edu/abs/1984A&A...134..189M}
}

@ARTICLE{Matteucci-2001,
       author = {{Matteucci}, Francesca and {Recchi}, Simone},
        title = "{On the Typical Timescale for the Chemical Enrichment from Type Ia Supernovae in Galaxies}",
      journal = {\apj},
         year = 2001,
        month = sep,
       volume = {558},
       number = {1},
        pages = {351-358},
          doi = {10.1086/322472},
archivePrefix = {arXiv},
       eprint = {astro-ph/0105074},
 primaryClass = {astro-ph},
       adsurl = {https://ui.adsabs.harvard.edu/abs/2001ApJ...558..351M}
}

@ARTICLE{Matsunaga-2020,
       author = {{Matsunaga}, Noriyuki and {Taniguchi}, Daisuke and {Jian}, Mingjie and {Ikeda}, Yuji and {Fukue}, Kei and {Kondo}, Sohei and {Hamano}, Satoshi and {Kawakita}, Hideyo and {Kobayashi}, Naoto and {Otsubo}, Shogo and {Sameshima}, Hiroaki and {Takenaka}, Keiichi and {Tsujimoto}, Takuji and {Watase}, Ayaka and {Yasui}, Chikako and {Yoshikawa}, Tomohiro},
        title = "{Identification of Absorption Lines of Heavy Metals in the Wavelength Range 0.97-1.32 {\ensuremath{\mu}}m}",
      journal = {\apjs},
         year = 2020,
        month = jan,
       volume = {246},
       number = {1},
          eid = {10},
        pages = {10},
          doi = {10.3847/1538-4365/ab5c25},
archivePrefix = {arXiv},
       eprint = {1911.11277},
 primaryClass = {astro-ph.SR},
       adsurl = {https://ui.adsabs.harvard.edu/abs/2020ApJS..246...10M}
}

@ARTICLE{Meiksin-2001,
       author = {{Meiksin}, Avery and {Bryan}, Greg and {Machacek}, Marie},
        title = "{Hydrodynamical simulations of the Ly{\ensuremath{\alpha}} forest: data comparisons}",
      journal = {\mnras},
         year = 2001,
        month = oct,
       volume = {327},
       number = {1},
        pages = {296-322},
          doi = {10.1046/j.1365-8711.2001.04719.x},
archivePrefix = {arXiv},
       eprint = {astro-ph/0102367},
 primaryClass = {astro-ph},
       adsurl = {https://ui.adsabs.harvard.edu/abs/2001MNRAS.327..296M}
}

@ARTICLE{Melendez-1999,
       author = {{Mel{\'e}ndez}, Jorge and {Barbuy}, Beatriz},
        title = "{Oscillator Strengths and Damping Constants for Atomic Lines in the J and H Bands}",
      journal = {\apjs},
         year = 1999,
        month = oct,
       volume = {124},
       number = {2},
        pages = {527-546},
          doi = {10.1086/313261},
archivePrefix = {arXiv},
       eprint = {astro-ph/9908296},
 primaryClass = {astro-ph},
       adsurl = {https://ui.adsabs.harvard.edu/abs/1999ApJS..124..527M}
}

@ARTICLE{Melendez-1999b,
       author = {{Mel{\'e}ndez}, Jorge},
        title = "{Mni hyperfine structure in the near-infrared}",
      journal = {\mnras},
         year = 1999,
        month = jul,
       volume = {307},
       number = {1},
        pages = {197-202},
          doi = {10.1046/j.1365-8711.1999.02674.x},
archivePrefix = {arXiv},
       eprint = {astro-ph/9906180},
 primaryClass = {astro-ph},
       adsurl = {https://ui.adsabs.harvard.edu/abs/1999MNRAS.307..197M}
}

@ARTICLE{Melendez-2009,
       author = {{Mel{\'e}ndez}, J. and {Asplund}, M. and {Gustafsson}, B. and {Yong}, D.},
        title = "{The Peculiar Solar Composition and Its Possible Relation to Planet Formation}",
      journal = {\apjl},
         year = 2009,
        month = oct,
       volume = {704},
       number = {1},
        pages = {L66-L70},
          doi = {10.1088/0004-637X/704/1/L66},
archivePrefix = {arXiv},
       eprint = {0909.2299},
 primaryClass = {astro-ph.SR},
       adsurl = {https://ui.adsabs.harvard.edu/abs/2009ApJ...704L..66M}
}

@ARTICLE{Meszaros-2012,
       author = {{M{\'e}sz{\'a}ros}, Sz. and {Allende Prieto}, C. and {Edvardsson}, B. and {Castelli}, F. and {Garc{\'\i}a P{\'e}rez}, A.~E. and {Gustafsson}, B. and {Majewski}, S.~R. and {Plez}, B. and {Schiavon}, R. and {Shetrone}, M. and {de Vicente}, A.},
        title = "{New ATLAS9 and MARCS Model Atmosphere Grids for the Apache Point Observatory Galactic Evolution Experiment (APOGEE)}",
      journal = {\aj},
         year = 2012,
        month = oct,
       volume = {144},
       number = {4},
          eid = {120},
        pages = {120},
          doi = {10.1088/0004-6256/144/4/120},
archivePrefix = {arXiv},
       eprint = {1208.1916},
 primaryClass = {astro-ph.SR},
       adsurl = {https://ui.adsabs.harvard.edu/abs/2012AJ....144..120M}
}

@ARTICLE{Nandakumar-2022,
       author = {{Nandakumar}, G. and {Ryde}, N. and {Montelius}, M. and {Thorsbro}, B. and {J{\"o}nsson}, H. and {Mace}, G.},
        title = "{The Galactic chemical evolution of phosphorus observed with IGRINS}",
      journal = {\aap},
         year = 2022,
        month = dec,
       volume = {668},
          eid = {A88},
        pages = {A88},
          doi = {10.1051/0004-6361/202244724},
archivePrefix = {arXiv},
       eprint = {2210.04940},
 primaryClass = {astro-ph.SR},
       adsurl = {https://ui.adsabs.harvard.edu/abs/2022A&A...668A..88N}
}

@ARTICLE{Nissen-2015,
       author = {{Nissen}, P.~E.},
        title = "{High-precision abundances of elements in solar twin stars. Trends with stellar age and elemental condensation temperature}",
      journal = {\aap},
         year = 2015,
        month = jul,
       volume = {579},
          eid = {A52},
        pages = {A52},
          doi = {10.1051/0004-6361/201526269},
archivePrefix = {arXiv},
       eprint = {1504.07598},
 primaryClass = {astro-ph.SR},
       adsurl = {https://ui.adsabs.harvard.edu/abs/2015A&A...579A..52N}
}

@ARTICLE{Nissen-2020,
       author = {{Nissen}, P.~E. and {Christensen-Dalsgaard}, J. and {Mosumgaard}, J.~R. and {Silva Aguirre}, V. and {Spitoni}, E. and {Verma}, K.},
        title = "{High-precision abundances of elements in solar-type stars. Evidence of two distinct sequences in abundance-age relations}",
      journal = {\aap},
         year = 2020,
        month = aug,
       volume = {640},
          eid = {A81},
        pages = {A81},
          doi = {10.1051/0004-6361/202038300},
archivePrefix = {arXiv},
       eprint = {2006.06013},
 primaryClass = {astro-ph.SR},
       adsurl = {https://ui.adsabs.harvard.edu/abs/2020A&A...640A..81N}
}

@INPROCEEDINGS{Otsubo-2024,
       author = {{Otsubo}, Shogo and {Sarugaku}, Yuki and {Takeuchi}, Tomomi and {Ikeda}, Yuji and {Matsunaga}, Noriyuki and {McWilliam}, Andrew and {Hull}, Charlie and {Yoshikawa}, Tomohiro and {Katoh}, Haruki and {Kondo}, Sohei and {Hamano}, Satoshi and {Taniguchi}, Daisuke and {Kawakita}, Hideyo},
        title = "{WINERED fully commissioned at the Magellan Clay Telescope}",
    booktitle = {Ground-based and Airborne Instrumentation for Astronomy X},
         year = 2024,
       editor = {{Bryant}, Julia J. and {Motohara}, Kentaro and {Vernet}, Jo{\"e}l. R.~D.},
       series = {Society of Photo-Optical Instrumentation Engineers (SPIE) Conference Series},
       volume = {13096},
        month = jul,
          eid = {1309631},
        pages = {1309631},
          doi = {10.1117/12.3018617},
       adsurl = {https://ui.adsabs.harvard.edu/abs/2024SPIE13096E..31O}
}

@ARTICLE{Ramirez-2009,
       author = {{Ram{\'\i}rez}, I. and {Mel{\'e}ndez}, J. and {Asplund}, M.},
        title = "{Accurate abundance patterns of solar twins and analogs. Does the anomalous solar chemical composition come from planet formation?}",
      journal = {\aap},
         year = 2009,
        month = dec,
       volume = {508},
       number = {1},
        pages = {L17-L20},
          doi = {10.1051/0004-6361/200913038},
       adsurl = {https://ui.adsabs.harvard.edu/abs/2009A&A...508L..17R}
}

@ARTICLE{Ramirez-2014a,
       author = {{Ram{\'\i}rez}, I. and {Mel{\'e}ndez}, J. and {Bean}, J. and {Asplund}, M. and {Bedell}, M. and {Monroe}, T. and {Casagrande}, L. and {Schirbel}, L. and {Dreizler}, S. and {Teske}, J. and {Tucci Maia}, M. and {Alves-Brito}, A. and {Baumann}, P.},
        title = "{The Solar Twin Planet Search. I. Fundamental parameters of the stellar sample}",
      journal = {\aap},
         year = 2014,
        month = dec,
       volume = {572},
          eid = {A48},
        pages = {A48},
          doi = {10.1051/0004-6361/201424244},
archivePrefix = {arXiv},
       eprint = {1408.4130},
 primaryClass = {astro-ph.SR},
       adsurl = {https://ui.adsabs.harvard.edu/abs/2014A&A...572A..48R}
}

@ARTICLE{Ratcliffe-2024,
       author = {{Ratcliffe}, Bridget and {Minchev}, Ivan and {Cescutti}, Gabriele and {Spitoni}, Emanuele and {J{\"o}nsson}, Henrik and {Anders}, Friedrich and {Queiroz}, Anna and {Steinmetz}, Matthias},
        title = "{Chemical clocks and their time zones: understanding the [s/Mg]-age relation with birth radii}",
      journal = {\mnras},
         year = 2024,
        month = feb,
       volume = {528},
       number = {2},
        pages = {3464-3472},
          doi = {10.1093/mnras/stae226},
archivePrefix = {arXiv},
       eprint = {2307.11159},
 primaryClass = {astro-ph.GA},
       adsurl = {https://ui.adsabs.harvard.edu/abs/2024MNRAS.528.3464R}
}

@ARTICLE{Roederer-2014,
       author = {{Roederer}, Ian U. and {Jacobson}, Heather R. and {Thanathibodee}, Thanawuth and {Frebel}, Anna and {Toller}, Elizabeth},
        title = "{Detection of Neutral Phosphorus in the Near-ultraviolet Spectra of Late-type Stars}",
      journal = {\apj},
         year = 2014,
        month = dec,
       volume = {797},
       number = {1},
          eid = {69},
        pages = {69},
          doi = {10.1088/0004-637X/797/1/69},
archivePrefix = {arXiv},
       eprint = {1410.8539},
 primaryClass = {astro-ph.SR},
       adsurl = {https://ui.adsabs.harvard.edu/abs/2014ApJ...797...69R}
}

@ARTICLE{Ryabchikova-2015,
doi = {10.1088/0031-8949/90/5/054005},
url = {https://dx.doi.org/10.1088/0031-8949/90/5/054005},
year = {2015},
month = {apr},
publisher = {IOP Publishing},
volume = {90},
number = {5},
pages = {054005},
author = {T Ryabchikova and N Piskunov and R L Kurucz and H C Stempels and U Heiter and Yu Pakhomov and P S Barklem},
title = {A major upgrade of the VALD database},
journal = {Physica Scripta}
}

@ARTICLE{Schonebeck-2014,
       author = {{Sch{\"o}nebeck}, Frederik and {Puzia}, Thomas H. and {Pasquali}, Anna and {Grebel}, Eva K. and {Kissler-Patig}, Markus and {Kuntschner}, Harald and {Lyubenova}, Mariya and {Perina}, Sibilla},
        title = "{The Panchromatic High-Resolution Spectroscopic Survey of Local Group Star Clusters. I. General data reduction procedures for the VLT/X-shooter UVB and VIS arm}",
      journal = {\aap},
         year = 2014,
        month = dec,
       volume = {572},
          eid = {A13},
        pages = {A13},
          doi = {10.1051/0004-6361/201424196},
archivePrefix = {arXiv},
       eprint = {1409.4663},
 primaryClass = {astro-ph.IM},
       adsurl = {https://ui.adsabs.harvard.edu/abs/2014A&A...572A..13S}
}

@ARTICLE{Shejeelammal-2024,
       author = {{Shejeelammal}, J. and {Mel{\'e}ndez}, Jorge and {Rathsam}, Anne and {Martos}, Giulia},
        title = "{The [Y/Mg] chemical clock in the Galactic disk: The influence of metallicity and the Galactic population in the solar neighbourhood}",
      journal = {\aap},
         year = 2024,
        month = oct,
       volume = {690},
          eid = {A107},
        pages = {A107},
          doi = {10.1051/0004-6361/202449669},
archivePrefix = {arXiv},
       eprint = {2407.07283},
 primaryClass = {astro-ph.GA},
       adsurl = {https://ui.adsabs.harvard.edu/abs/2024A&A...690A.107S}
}

@ARTICLE{Sneden-1973,
       author = {{Sneden}, C.},
        title = "{The nitrogen abundance of the very metal-poor star HD 122563.}",
      journal = {\apj},
         year = 1973,
        month = sep,
       volume = {184},
        pages = {839},
          doi = {10.1086/152374},
       adsurl = {https://ui.adsabs.harvard.edu/abs/1973ApJ...184..839S}
}

@ARTICLE{Spina-2018,
       author = {{Spina}, Lorenzo and {Mel{\'e}ndez}, Jorge and {Karakas}, Amanda I. and {dos Santos}, Leonardo and {Bedell}, Megan and {Asplund}, Martin and {Ram{\'\i}rez}, Ivan and {Yong}, David and {Alves-Brito}, Alan and {Bean}, Jacob L. and {Dreizler}, Stefan},
        title = "{The temporal evolution of neutron-capture elements in the Galactic discs}",
      journal = {\mnras},
         year = 2018,
        month = feb,
       volume = {474},
       number = {2},
        pages = {2580-2593},
          doi = {10.1093/mnras/stx2938},
archivePrefix = {arXiv},
       eprint = {1711.03643},
 primaryClass = {astro-ph.SR},
       adsurl = {https://ui.adsabs.harvard.edu/abs/2018MNRAS.474.2580S}
}

@ARTICLE{Takeda-2024,
       author = {{Takeda}, Y.},
        title = "{Phosphorus Abundances of B-Type Stars in the Solar Neighborhood}",
      journal = {\actaa},
         year = 2024,
        month = sep,
       volume = {74},
       number = {1},
        pages = {43-67},
          doi = {10.32023/0001-5237/74.1.3},
archivePrefix = {arXiv},
       eprint = {2409.02742},
 primaryClass = {astro-ph.SR},
       adsurl = {https://ui.adsabs.harvard.edu/abs/2024AcA....74...43T}
}

@ARTICLE{Taniguchi-2025,
       author = {{Taniguchi}, Daisuke and {Matsunaga}, Noriyuki and {Kobayashi}, Naoto and {Jian}, Mingjie and {Thorsbro}, Brian and {Fukue}, Kei and {Hamano}, Satoshi and {Ikeda}, Yuji and {Kawakita}, Hideyo and {Kondo}, Sohei and {Otsubo}, Shogo and {Sameshima}, Hiroaki and {Tsujimoto}, Takuji and {Yasui}, Chikako},
        title = "{MAGIS (Measuring Abundances of red super Giants with Infrared Spectroscopy) project: I. Establishment of an abundance analysis procedure for red supergiants and its evaluation with nearby stars}",
      journal = {\aap},
         year = 2025,
        month = jan,
       volume = {693},
          eid = {A163},
        pages = {A163},
          doi = {10.1051/0004-6361/202452392},
archivePrefix = {arXiv},
       eprint = {2501.10502},
 primaryClass = {astro-ph.GA},
       adsurl = {https://ui.adsabs.harvard.edu/abs/2025A&A...693A.163T}
}

@ARTICLE{Taniguchi-2026,
       author = {{Taniguchi}, Daisuke and {de Laverny}, Patrick and {Recio-Blanco}, Alejandra and {Tsujimoto}, Takuji and {Palicio}, Pedro A.},
        title = "{Solar twins in Gaia DR3 GSP-Spec: I. Building a large catalog of solar twins with ages}",
      journal = {\aap},
         year = 2026,
        month = mar,
       volume = {707},
          eid = {A260},
        pages = {A260},
          doi = {10.1051/0004-6361/202658913},
archivePrefix = {arXiv},
       eprint = {2601.15387},
 primaryClass = {astro-ph.SR},
       adsurl = {https://ui.adsabs.harvard.edu/abs/2026A&A...707A.260T}
}

@ARTICLE{Tsujimoto-2021,
       author = {{Tsujimoto}, Takuji},
        title = "{Two Sites of r-process Production Assessed on the Basis of the Age-tagged Abundances of Solar Twins}",
      journal = {\apjl},
         year = 2021,
        month = oct,
       volume = {920},
       number = {2},
          eid = {L32},
        pages = {L32},
          doi = {10.3847/2041-8213/ac2c75},
archivePrefix = {arXiv},
       eprint = {2110.02261},
 primaryClass = {astro-ph.GA},
       adsurl = {https://ui.adsabs.harvard.edu/abs/2021ApJ...920L..32T}
}

@ARTICLE{Walsen-2024,
       author = {{Walsen}, Kurt and {Jofr{\'e}}, Paula and {Buder}, Sven and {Yaxley}, Keaghan and {Das}, Payel and {Yates}, Robert M. and {Hua}, Xia and {Signor}, Theosamuele and {Eldridge}, Camilla and {Rojas-Arriagada}, Alvaro and {Tissera}, Patricia B. and {Johnston}, Evelyn and {Aguilera-G{\'o}mez}, Claudia and {Zoccali}, Manuela and {Gilmore}, Gerry and {Foley}, Robert},
        title = "{Assembling a high-precision abundance catalogue of solar twins in GALAH for phylogenetic studies}",
      journal = {\mnras},
         year = 2024,
        month = apr,
       volume = {529},
       number = {3},
        pages = {2946-2966},
          doi = {10.1093/mnras/stae280},
       adsurl = {https://ui.adsabs.harvard.edu/abs/2024MNRAS.529.2946W}
}
\bibliographystyle{aasjournal}



\end{document}